\documentclass[aps,prx,twocolumn,superscriptaddress,showkeys,floatfix]{revtex4-1}
\usepackage{color}
\usepackage{amsmath}
\usepackage{amsfonts}
\usepackage{dsfont}
\usepackage{graphicx}
\usepackage{subfigure}
\usepackage{transparent}
\definecolor{darkgreen}{rgb}{0,0.4,0}
\ifx\pdftexversion\undefined
\usepackage[dvips]{hyperref}
\else
\usepackage{hyperref}
\fi
\hypersetup{
  colorlinks = true, linkcolor = magenta,  citecolor=darkgreen
}
\definecolor{darkgreen}{rgb}{0,0.4,0}
\renewcommand{\phi}{\varphi}

\renewcommand{\theta}{\vartheta}

\begin{document}

\title{Engineering matter interactions using squeezed vacuum}

\author{Sina Zeytino$\breve{\mathrm{g}}$lu}
\affiliation{Institute for Theoretical Physics, ETH Zurich, CH-8093 Z{\"u}rich, Switzerland.}
\affiliation{Institute of Quantum Electronics, ETH Zurich, CH-8093 Z{\"u}rich, Switzerland.}

\author{Ata\c c $\dot{\mathrm{I}}$mamo$\breve{\mathrm{g}}$lu}
\affiliation{Institute of Quantum Electronics, ETH Zurich, CH-8093 Z{\"u}rich, Switzerland.}

\author{Sebastian Huber}
\affiliation{Institute for Theoretical Physics, ETH Zurich, CH-8093 Z{\"u}rich, Switzerland.}

\date{\today}

%TC:ignore 
\begin{abstract} 

Virtually all interactions that are relevant for atomic and condensed matter physics are mediated by quantum fluctuations
of the electromagnetic field vacuum. 
Consequently, controlling the vacuum fluctuations can be used to engineer the strength and the range of interactions.
Recent experiments have used this premise to demonstrate novel quantum phases or entangling gates by embedding 
electric dipoles in photonic cavities or waveguides, which modify the electromagnetic fluctuations.
Here, we show theoretically that the enhanced fluctuations in the anti-squeezed quadrature of a squeezed vacuum state
allows for engineering interactions between electric dipoles without the need for a photonic structure. 
Thus, the strength and range of the interactions can be engineered in a time-dependent way by changing the
spatial profile of the squeezed vacuum in a travelling-wave geometry, which also allows the implementation
of chiral dissipative interactions. Using experimentally realized squeezing parameters 
and including realistic losses, we predict single atom cooperativities $C$ of up to 10 for the squeezed vacuum enhanced
interactions.

%
%Virtually all interactions that are relevant for atomic and condensed matter physics are mediated
%by the vacuum fluctuations of the electromagnetic field. 
%
%Thus, these interactions can be enhanced 
%by modifying the vacuum fluctuation spectrum. Here, we present vacuum anti-squeezing 
%as a way of enhancing dipolar interactions by modifying vacuum fluctuations. 
%
%The range of the resulting inter-atomic coupling can be engineered by dynamically changing the spatial profile 
%of the squeezed vacuum in a traveling wave geometry. 
%
\end{abstract}
%TC:endignore 
\maketitle

\section{Introduction}
%\begin{figure}
%\includegraphics{test}
%\end{figure}
The accurate control of quantum degrees of freedom lies at the heart of many fields of research.
The applications of quantum control range from digital quantum information processing
\cite{nielsen2010quantum, Cirac1995, Wallraff2004, Majer2007, Imamoglu1999, Saffman2010}, 
where quantum correlations are harnessed by controlling a few, well-isolated quantum bits,
all the way to the manipulation of the collective degrees of freedom in complex many-body systems
\cite{Blatt2012, Landig2016, Schreiber2016,Labuhn2016}. 
Interactions play a key role in almost all quantum control schemes as a means for entanglement-generation
between a few or macroscopically many quantum degrees of freedom. 
In turn, interactions in practically all physical implementations are mediated by fluctuations
 of the electromagnetic vacuum.
Thus, it is highly desired to have an experimental framework, where the electromagnetic vacuum can be modified
both spatially and temporally in order to realise quantum systems with
non-trivial correlations.
Here, we propose to use anti-squeezing of the electromagnetic vacuum to achieve these design principles. 
We focus on the dispersive as well as dissipative dipole-dipole interactions that are
mediated by the enhanced fluctuations of the anti-squeezed vacuum quadrature.

The main object that describes the interactions between electromagnetic vacuum 
modes and a dipole transition is the spectral density $J(\omega)$ defined as
$J(\omega)~=~\sum_{l}|\eta_{l}|^2\delta(\omega - \omega_l)$,
where $l$ labels the electromagnetic vacuum states and $\eta_{l}$  is the dipole coupling strength
to the vacuum state. Thus, the $J(\omega)$ encodes the information of both the density 
of states of modes at a frequency $\omega$ as well as the coupling strength of the 
atomic dipole to these modes. Moreover, as we shall show below, for a non-interacting bath, the effective
interactions between two dipoles at positions $r$ and $r'$ is given by 
$\int_{-\infty}^{\infty} d\omega f_{\omega}(r,r')J(\omega)2\omega/(\omega^2_{\mathrm{dip}}-\omega^2)$,
where $f_{\omega}(r,r')$ encodes the spatial overlap between a given set of
the electromagnetic field modes and the dipoles, and $\omega$ is in a suitable rotating frame.

%$\int_{0}^{\infty} d\omega \bar{J}(\omega)2\omega/(\omega^2_{\mathrm{dip}}-\omega^2)$,
%where $\omega_{\mathrm{dip}}$ is the dipole transition frequency in a suitable rotating frame
%and $\bar{J}(\omega)$ is the spectral density in the lab frame.

\begin{figure}[h!]
\includegraphics{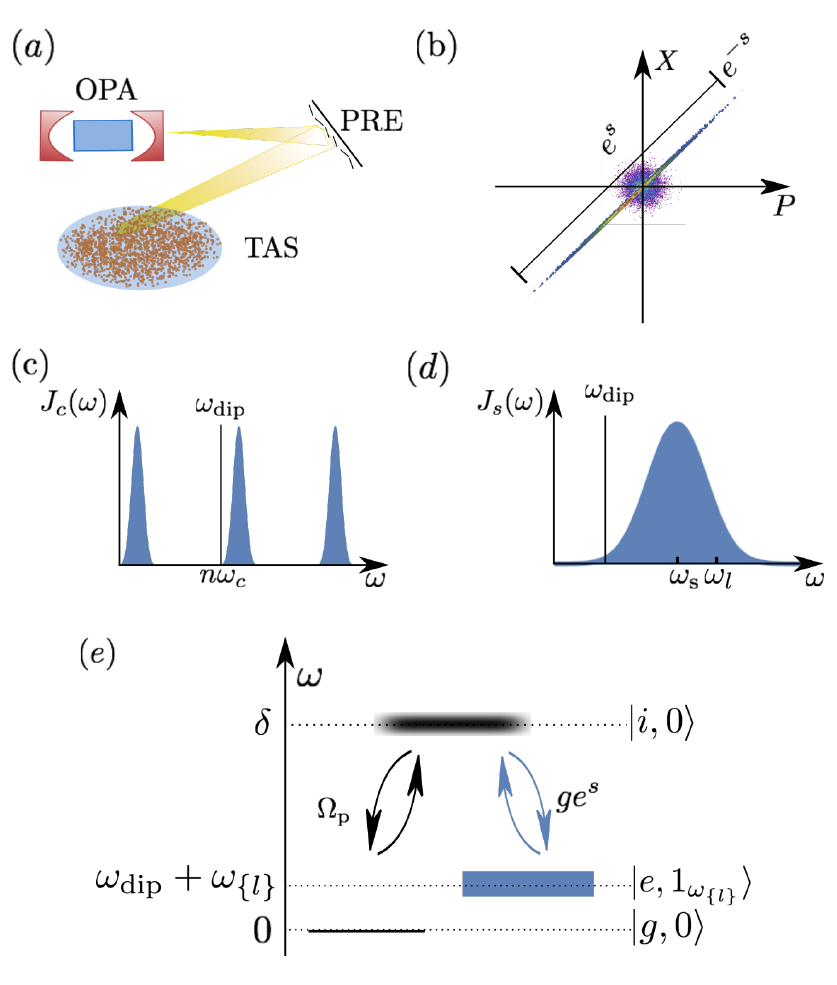}
\caption{``Color" (a) Experimental setup. The squeezed vacuum is generated via the non-linear processes in the 
optical parametric amplifier (OPA) operated below threshold. The squeezed vacuum is then reflected by a 
programmable reflective element (PRE) such as a spatial light modulator (SLM) or a digital mirror device (DMD) and focused 
to a diffraction limited pattern on the trapped atomic system (TAS). (b) The effect of squeezing in the phase space.
Squeezing results in enhanced fluctuations in one quadrature while reducing the fluctuations in the orthogonal 
quadrature. (c) The spectral density $J_{c}(\omega)$ for a dipole placed within a cavity. The boundary conditions imposed
by the cavity mirrors result in a highly modified density of states within the cavity. The change in
density of states modifies the $J_{c}(\omega)$ and results in the phenomenology of cQED. (d) The spectral density 
for a dipole coupled to the anti-squeezed quadrature of the squeezed vacuum beam with a carrier frequency
$\omega_{\mathrm{s}}$. The modes within the bandwidth of the squeezed vacuum are labeled by $l\in\{-L,L\}$.
(e) A typical application of squeezed vacuum with squeezing parameter $s$ in a stimulated 
Raman transition that requires a coherent drive with strength
$\Omega_\mathrm{p}$. The level structure of the quantum degrees of freedom are depicted in the rotating frame 
of the coherent pump. The labels $g$ and $e$ denote the two metastable qubit states separated by $\omega_{\mathrm{dip}}$
and $i$ denotes the short lived intermediate state whose detuning from the coherent pump is given by $\delta$.
All states are shown in a tensor product form (i.e., $|\cdots\rangle \otimes |\cdots\rangle$) with the state of the squeezed electromagnetic modes.
The effective coupling strength between the ground and the excited states is $ g e^{s} \Omega_p/\delta$.
We note that the coupling to the dipole operator Eq. ($\ref{eq:BEC}$) is realized when the Raman
transition brings the system back to $|g\rangle$, whereas coupling to the operator in Eq. ($\ref{eq:spin}$) 
necessitates a balanced Raman scheme.}
\label{fig:ExpSet}
\end{figure}

In order to understand the role of squeezed vacuum in engineering dipolar interactions,
it is helpful to discuss the modification of $J(\omega)$ in cavity quantum electrodynamics 
(cQED) \cite{Haroche2006exploring}, where it drastically changes the dispersive and dissipative dynamics of
the dipoles placed within the cavity. The density of states within a high-finesse cavity is modified
due to the interference of multiply scattered photons. 
As a result, $J_{\mathrm{c}}(\omega)$ has Lorentzian peaks at integer multiples of the cavity resonance 
frequencies $\omega_{\mathrm{c}}$, while the fluctuations are suppressed over the free spectral range between the 
resonance frequencies [see Fig. \ref{fig:ExpSet} (c)].

Many aspects of the dissipative and dispersive dynamics of an electric dipole placed inside the cavity can be
understood with the help of  $J_{\mathrm{c}}(\omega)$. 
The condition $J_{\mathrm{c}}(\omega_{\mathrm{dip}}) > J_{\mathrm{free}}(\omega_{\mathrm{dip}})$ 
results in the celebrated Purcell enhancement  \cite{Purcell1946} of the decay rate of an excited dipole.
Moreover, any asymmetry of $J_{\mathrm{c}}(\omega)$ with respect to $\omega_{\mathrm{dip}}$ results
in a cavity contribution to the Lamb shift \cite{Lamb1947,Haroche2006exploring,Wallraff2004}. 
When $J_{\mathrm{c}}(\omega)$ is non-constant in a frequency window of order $J_{\mathrm{c}}(\omega)/\hbar$, the cavity 
vacuum should be considered a non-Markovian bath, resulting in reversible processes such as the 
vacuum Rabi oscillations \cite{Haroche2006exploring}. \textit{Most importantly} for the purposes of this work, 
when multiple dipoles are placed within the cavity, the dissipative (Purcell) and dispersive (Lamb) effects 
associated with a single atom translate to dissipative and dispersive interactions between the atoms. 

A squeezed vacuum state is characterised
by the absence of the quadrature rotation [$U(1)$] symmetry
of the vacuum fluctuations [see Fig. \ref{fig:ExpSet}(b)]. However, the two major orthogonal quadratures of the squeezed vacuum 
remain conjugate variables such that the enhanced fluctuations in one quadrature is accompanied by
reduced fluctuations in the other. Thus, in the phase space representation, the squeezed vacuum state is given by 
an ellipse centered at the origin. 

The squeezed vacuum spectral density 
$J_{\mathrm{sq}}$ resembles $J_{\mathrm{c}}$ in some important aspects [compare Fig.~\ref{fig:ExpSet}~(c)~and~(d)].
The vacuum squeezing modifies the spectral function $J_{\mathrm{free}}(\omega)$ not by changing the density of
states, but rather by increasing the dipole interaction strength $\eta_{l}$. 
To see this, note that the interaction strength $\eta_l^{0}$ between the anti-squeezed quadrature $\hat{X}_l^{0}$ and the atomic 
dipole transition increases exponentially with the squeezing parameter $s$ [see Fig.~\ref{fig:ExpSet}~(b)]
\cite{gardiner2004quantum}
\begin{align}
H_{dip}  = \eta_l^{0}\hat{O}\hat{X}_l^{0} \rightarrow H_{dip}^{sq}  = \eta_l^{0} e^{s}\hat{O}\hat{X}_l,
\label{eq:secret}
\end{align} 
where  $\hat{O}$ is the atomic dipole operator and we defined a new quadrature operator $X_{l}$ whose 
fluctuations are normalized. Such an enhancement of interactions between electronic transitions and anti-squeezed bosons
has been discussed previously in superconductivity \cite{Hakioglu1997,Cotlet2016, Knap2016}, and optomechanics \cite{Lemonde2016,Lu2015}.
Furthermore, $J_{\mathrm{sq}}$ is similar to $J_{\mathrm{c}}$ in that both can depend strongly
on frequency, since the anti-squeezing is typically realised over a narrow bandwidth of a few MHz \cite{Mehmet2011}.
A schematic for $J_{\mathrm{sq}}$ for squeezed vacuum with a squeezing frequency
$\omega_s$ is shown in Fig.~$\ref{fig:ExpSet} (d)$. 

It is important to note that most of the pioneering studies  \cite{Gardiner1986, Carmichael1987, Walls1983, Dalton1999} 
considered the effect of squeezed vacuum on the dissipative properties of a single two-level system. On the other hand,
with our proposal, we aim to demonstrate that the squeezed vacuum can be used as a resource for engineering dispersive as well as
dissipative interactions between any number of pairs of emitters. Moreover, instead of focusing on the reduced
fluctuations in the \textit{squeezed} vacuum quadrature, we seek to take advantage of the enhanced fluctuations 
in the \textit{anti-squeezed} quadrature.

%It is also important to note that unlike the proposals that
%employ squeezed vacuum for precision measurements, which focus on taking advantage of the reduced
%fluctuations in the squeezed quadrature, we focus on the enhanced fluctuations in the anti-squeezed quadrature.

We argue that squeezed vacuum, when used as a mediator of interactions provides versatile control
knobs for artificial quantum many body systems. Most importantly, the squeezed vacuum can be used 
in a \textit{traveling-wave} geometry, allowing the spatial profile of the squeezed vacuum modes
to be shaped independently of the atomic system. This is in drastic contrast to the 
setups usually employed in cQED and trapped ions where the spatial profile of the 
virtual bosons that mediate interactions are determined by the dielectric material which confines 
the vacuum fluctuations. Because the atoms have to be trapped in the close vicinity of the confining
dielectric, the interaction mediating bosons cannot be modified independently of the atoms. 
%
%In these systems, the atoms should be trapped in close proximity to the 
%dielectric material, whose characteristics cannot be changed \textit{in-situ}.

More specifically, the spatial dependence of the squeezed vacuum modes in the traveling wave 
geometry can be determined with diffraction limited resolution by using a programmable reflective 
element such as a spatial light modulator (SLM) or a digital mirror device (DMD). Moreover, thanks to the programmability
of SLM and DMD's, the spatial profile of the squeezed vacuum modes can be modified in a  \textit{time-dependent} fashion (up to frame rates 
of 100 kHz for DMD's), allowing for the implementation of time-averaged and time-dependent interaction Hamiltonians.
%In this sense, the pulses of squeezed vacuum can be thought of as flying cavities: 
%wave packets of vacuum with modified fluctuations. 
In this work, we emphasize that an
important advantage of having such dynamical control is the possibilty of implementing translationally invariant
interactions with arbitrary spatial dependence as well as disordered interactions with arbitrary disorder correlations \cite{Cirac2012}.
We also note that the traveling wave geometry of the squeezed vacuum setup can be utilised to implement
chiral dissipative interactions \cite{Lodahl2017, Ramos2014}.

The dynamical spatial tunability that our proposal offers comes at the price
of diffractive losses that occur along the optical path of the squeezed vacuum beam. However, 
it is important to note that unlike the fluctuations
in the squeezed quadrature, which are infamously fragile against losses \cite{Mehmet2011},
the enhanced fluctuations in the anti-squeezed quadrature decrease only linearly with losses in the limit of
large squeezing \cite{Tomaru2006}. Thus, the modification of the spatial profile of the squeezed vacuum modes 
can be performed without a drastic loss of anti-squeezing in the vacuum fluctuations. To illustrate this stability, 
we show that single atom cooperativities of order unity are achievable including realistic losses.
% \textcolor{red}{
%Specifically, a squeezed vacuum beam focused on the emitter in a diffraction limited fashion,
%a single atom cooperativity of order 1 can be achieved for up to 80 \% diffractive losses and a degree of squeezing 
%that is experimentally realized  \cite{Mehmet2011} in the optical regime. The cooperativity for the same setup with 
%50 \% losses is about 6.}

In the following, we first briefly review the generation of multimode squeezed vacuum in a nonlinear optical cavity. 
Then, we derive the terms describing the effective dispersive and dissipative interactions mediated by the 
fluctuations in the squeezed vacuum. We also show how interactions with an arbitrary spatial dependence can be implemented using DMD's. 
To assess the experimental feasibility of the squeezed vacuum setup, we estimate the single atom cooperativity 
that can be achieved using experimentally demonstrated squeezing parameters. Lastly, we discuss some applications
of the squeezed vacuum setup in realising exotic quantum phases. 

% Then we discuss two models which have been previously
%discussed in the multimode cQED framework. Following the previous work in Ref. \cite{Gopalakrishnan2011, Gopalakrishnan2010} 
%we consider two atomic models: one describing two-level bosonic atoms in a shallow trap and the other describing
%three-level atoms in $\Lambda$ configuration in a deep optical lattice. Finally, we define and estimate the cooperativity for the
%proposed squeezed vacuum setting, and discuss the experimental applications including finite range and disordered interactions.
%
%\textcolor{red}{put already the enhancement of the dipole coupling here. Also try to state that the anti-squeezed quadrature 
%is more robust than the squeezed quadrature as this is what would ring in peoples minds.} 

\section{Squeezed vacuum generation with an optical parametric amplifier}
\label{sec:OPO}

%Squeezed vacuum states resulting from the nonlinear optical response of polarisable media \cite{Bloembergen1996nonlinear}
%Squeezed vacuum state can be identified by its zero mean value for the photon density and the non-vanishing 
%anomalous expectation(?How to make this more colloquial?) value for two particle excitations taken over an ensemble
%\cite{note that for any realisation of squeezed vacuum state the number of photons is a conserved quantity. Similarly to 
%the case of superconductivity, the anomalous expectation values are non-zero only when the expectation value is 
%averaged over ensembles.}. The anomalous expectation value of the two particle excitations signal the breaking 
%of the rotation symmetry in the quadrature basis. Indeed, the fluctuations of the squeezed vacuum along the two 
%principle quadrature axes, which are called the squeezed and the anti-squeezed quadrature are scaled by the 
%squeezing degree according to $e^{\pm s}$, respectively. 

The squeezed vacuum is routinely generated via nonlinear processes in polarizable media \cite{Bloembergen1996nonlinear}.
Here we consider an optical parametric amplifier (OPA) setup where the nonlinear medium is placed inside a 
high finesse cavity (see Fig. \ref{fig:ExpSet} (a)), and pumped by a coherent drive at twice the squeezing frequency $2\omega_s$. 
We assume that the OPA is operated
below the oscillation threshold, where the intracavity gain
is smaller than the losses, and the output of the non-linear process has a zero mean. 

Under these conditions, the creation(annihilation) operator $(\bar{a}^{\dagger}(\bar{a}))$
for the output fields of the OPA satisfies the equation
%Solving the Quantum Langevin equations for the  $\chi^{(2)}$ process for a one-sided cavity
%yields the following expression for the creation and annihilation operators of the output fields  $a^{\sigma}(\omega)$
\cite{gardiner2004quantum}:
\begin{align}
\bar{a}(\omega) &= u(\omega) a_{in} (\omega) - v(\omega) a_{in}^{\dagger}(2\omega_s-\omega),
\label{eq:SqueezTrans}
\end{align}
where the operators $a_{in}$ and $a^{\dagger}_{in}$
describe the quantum noise injected into the system from the environment, and ensure
that the output operators obey bosonic commutation relations \cite{Sachdev1984,Carmichael1999statistical}.
The complex coefficients $u(\omega)$ and $v(\omega)$ obey the bosonic normalisation condition
\begin{align}
\nonumber |u(\omega)|^2-|v(\omega)|^2 = 1,
\end{align}
and describe the squeezing in the output fields. In particular, they define the squeezing parameter $s(\omega)$ by the relation
\begin{align}
e^{\pm s(\omega) } \equiv |u(\omega)\pm v(\omega)|,
\end{align}
where we have set the squeezing angle to zero (see Appendix \ref{app:SqueezeSpec}).
Then, the fluctuations in the squeezed and anti-squeezed quadratures are scaled by $e^{\pm s(\omega)}$, respectively 
[see Fig. $\ref{fig:ExpSet}$ (b)]. 

Generically, the frequency bandwidth over which the squeezing degree $s(\omega)$ is non-zero is determined
by the lifetime of the intra-cavity photons of the OPA. Recently, anti-squeezing of approximately 20 dB \footnote{The enhancement
of vacuum fluctuations is expressed in decibels as $e^{(2s)}= 10^{X\mathrm{dB}/10}$. Thus, 20 dB anti-squeezing gives
and enhancement of factor 100 in the fluctuations.}
over a 10 MHz bandwidth at optical frequencies has been reported \cite{Mehmet2011, Vahlbruch2008}. 
It is important to note that the focus of these experiments was to achieve the largest 
squeezing possible, which is limited by detection efficiency as much as it is limited by the interaction strength 
inside the OPA. It is in principle possible to get larger anti-squeezing by driving the OPA closer to the oscillation
threshold.

\section{Atom-field coupling}
For the sake of generality, we assume linear coupling between 
a quadrature $\hat{X}$ of the squeezed vacuum modes and a Hermitian operator $\hat{O}$ describing atomic transitions.
In addition, we assume that all the dipole moments are aligned.
In the following, we allow the complex amplitudes $\tilde{\Xi}(r)$ of the squeezed vacuum modes to have arbitrary spatial
dependencies, since these amplitudes can be shaped using programmable reflective elements as shown in Fig. \ref{fig:ExpSet}(a).

The model Hamiltonian we use to describe the atom-field system is given by (see Appendix $\ref{app:Cascade}$)
\begin{widetext}
\begin{align}
\nonumber H/\hbar &=\int d^{2}r (H_{\mathrm{at}}(r) + H^{s}_{\mathrm{vac}}(r) + H_{\mathrm{dip}}(r))/\hbar\\
& =\int d^{2}r H_{at}(r)/\hbar + \left(\sum_{l}  \omega_l |\Xi_l |^{2}(r)\bar{a}^{\dagger}(\omega_l)\bar{a}(\omega_l) +  \hat{O}(r) \sum_{l} \eta e^{s(\omega_l)}\bar{X}_l(r) \right),
\label{eq:BosHam}
\end{align}
\end{widetext}
where $\hat{X}_l$ is the squeezed vacuum quadrature given by 
\begin{align}
\bar{X}_l (r)=  \tilde{\Xi}_{l}(r) \bar{a}(\omega_l)  + (\tilde{\Xi}_{l})^*(r) \bar{a}^{\dagger}(2\omega_s-\omega_l), 
\end{align}
with $\tilde{\Xi}_{l}(r) = \Xi_{l}(r)e^{i\phi_l}$ determined by the squeezing transformation. The index
$l$ labels the modes within the squeezing
bandwidth and $\eta$ is the effective dipole interaction strength, which we assumed to be independent of $l$.
The coupling term between the squeezed vacuum quadrature and the atoms arises as 
a low energy description when the atoms are driven in a Raman scheme \cite{Ritsch2013, Dimer2007, Buchhold2013}. 
Thus, the time-independent Hamiltonian in Eq. ($\ref{eq:BosHam}$) is in the rotating frame that is 
determined by the frequency $\omega_{\mathrm{p}}$ of the coherent pump.
For instance the coupling strength for the scheme in Fig.~$\ref{fig:ExpSet}$~(e) is
$\eta = g \Omega_{\mathrm{p}}/\delta$. Here, $g$ is the bare dipole coupling strength to the electromagnetic vacuum,
$\Omega$ is the Rabi frequency due to the coupling to the coherent electromagnetic drive, and $\delta$ is the detuning between 
the coherent drive frequency and the dipole transition between the ground $|g\rangle$ and the intermediate $|i\rangle$ states.
In addition, we emphasize that the quadrature 
coupling is not essential for the enhancement of interactions. When a Jaynes-Cummings (JC) type interaction
is considered, the enhancement
of the interaction is simply reduced to half since $a(\omega_l) = \frac{1}{2}\left(\hat{X}_{l} + \hat{P}_l\right)$ (see Appendix~\ref{app:QuadJC}). A simple example of how
Eq.~($\ref{eq:BosHam}$) can emerge as an effective description of a driven condensate is given in Appendix $\ref{app:BoseEx}$.

\begin{figure}
\includegraphics{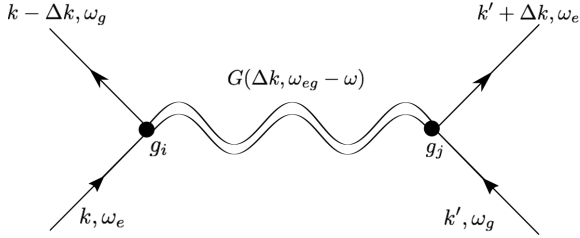}
\caption{The Feynman diagram describing the effective interactions between dipoles.}
\label{fig:Greens}
\end{figure}

Next, integrating out the squeezed vacuum modes,
we obtain pairwise coupling between all pairs of atomic operators $\hat{O}^{\dagger}_i \hat{O}_j$ for all atoms $i$ and $j$
which spatially overlap with the squeezed vacuum states.
The Hermitian part of the  effective interaction is given by

\begin{align}
\nonumber \tilde{H}_{int} &\approx \frac{1}{2}\iint d^2 r d^2 r' \int d\omega \frac{J(\omega)2\omega}{\omega^2_{\mathrm{dip}}-\omega^2}\tilde{\Xi}_\omega^*(r)\tilde{\Xi}_{\omega}(r') \hat{O}(r)\hat{O}(r')\\
&= \frac{1}{2} \sum_{l}\iint d^2 r d^2 r' \tilde{g}_l e^{2s(\omega_l)}\tilde{\Xi}_l^{*}(r)\tilde{\Xi}_{l}(r') \hat{O}(r)\hat{O}(r'),
\label{eq:DDInt} 
\end{align}
where $\tilde{g}_l \equiv 2\omega_l |\eta|^2/(\omega^2_{\mathrm{dip}} - \omega^2_l)$, and we have used
 $J(\omega)~=~\sum_{l}|\eta|^2 \delta(\omega-\omega_l)$. 
We note that to get the above expression, we have projected the interactions to the manifold with no excitation above the squeezed 
vacuum, ignoring the AC Stark shift terms.
The derivation of the effective interaction term using a projector method is given in Appendix \ref{app:Greens}.
Derivation of similar terms can be found in \cite{Ritsch2013, Buchhold2013, Douglas2015,Strack2011,Gopalakrishnan2011}.
%we have neglected the vacuum induced AC
%Stark shift terms as we are interested in dynamics where the occupation of the vacuum modes is negligible \footnote{
%These terms can become relevant if a coherent state spontaneously builds up in the vacuum modes. For instance, such 
%an effect is responsible for the chaotic behaviour of the single mode Dicke model near the phase transition. However, 
%as we are working with free space squeezed modes, the number of electromagnetic vacuum modes is large enough to 
%suppress the chaotic behaviour \cite{Tolkunov2007}}. 
%foxxr the condensate model. On the other hand, for the spin 1/2 model, we obtain
%\begin{align}
%\nonumber\tilde{H}&^{s}_{int}  
%%\sum_{l, \sigma = \pm}\sum_{i,j} \left(\Xi_l^{\sigma}(r_i)\right)^*\Xi_{l}^{\sigma}(r_j) \bigg(2(u_{l}v_{-l})\sigma^+(r_i)\sigma^+(r_j) \\
%%&+2(u^{*}_{l}v^{*}_{-l})\sigma^-(r_i)\sigma^-(r_j)+2(|u_l|^2 +|v_{l}|^{2})\sigma^+(r_i)\sigma^-(r_j)\bigg)\\
%&=  \sum_{l, \sigma= \pm}\sum_{i<j} \tilde{g}_l \left(\Xi_l^{\sigma}(r_i)\right)^*\Xi_{l}^{\sigma}(r_j) \tau_{ij}^{-}(l)\tau_{ij}^{+}(l) + h.c. \, .
%\label{eq:THeis} 
%\end{align}
%This approximation is valid when the atomic degrees of freedom evolve slowly compared to the 
%time scales of the squeezed vacuum modes. 
Because the induced interaction only depends on the spatial overlap between the atoms and the vacuum mode,
the interaction Hamiltonian in Eq. ($\ref{eq:DDInt}$) describes an infinite-range interaction as long as the vacuum 
mode is coherent between points $r$ and $r'$ (see Appendix \ref{app:Greens}). The coherence length of the squeezed vacuum mode 
is roughly given by the squeezing bandwidth. When the space-like separation between the 
dipoles is relevant, relativistic retardation applies to information transfer via such infinite-range interactions. 

The anti-Hermitian part of the effective interaction describes the correlated decay processes induced by real photons
emitted into the squeezed vacuum. The rate of correlated decay for two particles at $r$ and $r'$ is given by the Fermi's Golden rule
\begin{align}
\tilde{\gamma}~=~ \frac{2\pi}{\hbar} J_s (\overline{\omega}) =  \frac{2\pi}{\hbar} \sum_{l} \eta^2 \Xi^*_l(r)\Xi_l(r') e^{2s(\omega_l)}\delta(\bar{\omega}-\omega_l) ,
\end{align}
where $\overline{\omega}$ is the renormalised dipole transition frequency. 
Besides describing the correlated decay processes into the squeezed vacuum modes, the anti-Hermitian part of the 
effective interaction can also be used to increase the detection efficiency of light emitted from a single quantum emitter if the 
decay rate $\tilde{\gamma}$ is comparable to the decay rate into all the other unsqueezed vacuum modes \footnote{We note
that the enhanced fluctuations in the anti squeezed quadrature does not pose a problem for the detection scheme when the 
polarization of the squeezed vacuum modes are used appropriately. 
The light emitted from the atom into the squeezed vacuum can be measured 
independently from the squeezed vacuum by using the polarization of the squeezed vacuum modes \cite{Delteil2016}.
When the polarisation of the squeezed vacuum beam is not aligned with the polarisation of the light emitted by the 
atoms, the emitted light can be separated from the fluctuations in the anti-squeezed quadrature of the vacuum by 
the use of a polariser}. Moreover, a large cooperativity of the coupling to the squeezed vacuum modes
in a traveling-wave geometry entails that an emitter is more likely to emit photons in the propagation direction of the squeezed 
vacuum beam. In this respect, squeezed vacuum setup provides a natural framework for building chiral networks \cite{Lodahl2017, Ramos2014} 
(see Appendix~$\ref{app:Cascade}$).

In other words, while the Hermitian part of the effective interaction allows for engineering the Hamiltonian of the atomic system, the 
anti-Hermitian part allows for engineering the effective reservoir that is responsible for the dissipative dynamics.
The dissipative interactions can be implemented between arbitrarily distant dipoles as the strength of dissipative
interactions depend not on the squeezing bandwidth, but only on the squeezing parameter at the dipole transition 
frequency, as long as the squeezing bandwidth is larger than the linewidth of the transition.
 The situation where the squeezed reservoir results in entanglement between two distant qubits through dissipative 
 dynamics was discussed in 
 Ref. \cite{Kraus2004,Goldstein1996,Banarjee2010}.  

We note that the effective interaction between the atomic degrees of freedom can also be calculated in 
frequency space with the help of the Feynman diagram in Fig. $\ref{fig:Greens}$. This diagram has the obvious interpretation that the 
effective interactions between the atomic degrees of freedom are mediated by the dressed photons in the 
anti-squeezed quadrature. Consequently, what determines the strength and range of the effective dispersive and dissipative interactions 
are the fluctuations in the electromagnetic vacuum, which are described by dressed photon Green's function \cite{Kulkarni2013}.
Using the Green's function, we show in Appendix \ref{app:Greens}, that the enhanced quadrature fluctuations in a thermal or a Fock state do not lead 
to an enhancement in the effective coupling, while squeezed number and squeezed thermal states \cite{Kim1989} 
result in the same enhancement as the squeezed vacuum. 
%Moreover, we show that the enhancement of dipole interactions
%due to vacuum squeezing can be understood as a result of the energetic softening of the vacuum modes. 

Lastly, we discuss how the high tunability of the squeezed vacuum modes can be used to implement
translationally invariant finite-range interactions
with arbitrary spatial dependence, as well as disordered and time dependent interactions. 
Because the digital mirror device (DMD) provides a dynamic tuning knob on the spatial profile of the 
squeezed vacuum modes, it is possible to implement time-averaged Hamiltonians as long as the 
cycling rate $2\pi \tau_c^{-1}$ of the time-periodic Hamiltonian is much faster than the renormalised dynamics of the atomic system.
That is, we need 
\begin{align}
 |H_{at}| \tau_{c} \ll1.
 \end{align}
The cycling rate should also obey the adiabaticity condition with respect to the energy gap
  between states with no excitations and a single excitation above the squeezed vacuum.
 This energy gap can be increased by detuning the atomic transitions further away from the squeezed vacuum,
 which inevitably reduces the highest achievable effective interaction strength [ see Eq.~($\ref{eq:DDInt}$) ]
Experimentally, the lower limit of $|H_{at}|$ is determined by the lifetime of the atomic states, whereas the cycling rate can be up to 
100 kHz for DMD's. Remarkably, any translationally invariant interaction potential can be obtained by 
time-averaging over many speckled vacuum modes \cite{goodman2007speckle}. If each speckle realisation samples an ensemble 
characterized by a finite-range spatial autocorrelation function, then the time-averaged Hamiltonian effectively simulates
finite range interactions whose spatial dependence is given by the speckle autocorrelation function. That is,
the system evolves under an effective time(disorder)-averaged Hamiltonian \cite{Haeberlen1968}. 
\begin{align}
\tilde{H}_{\mathrm{int}}^{tav} \approx  \sum_{\mathbf{l}} \sum_{i<j}\tilde{g}_l e^{2s(\omega_l)}\overline{\langle \Xi_l^*(x_i-x_j)\Xi_l(0) \rangle} \hat{O}^{\dagger}(x_i)\hat{O}(x_j),
\end{align}
where $\overline{\langle \Xi_l^*(x_i-x_j)\Xi_l(0) \rangle}$ denotes the time(disorder) averaged autocorrelation function.
We note that this strategy can be used to implement both dissipative and dispersive finite range interactions
The number of disorder samples can be further increased by splitting the squeezed vacuum 
beam and propagating it over different DMD's.
On the other hand, in the regime where the time-averaging condition is not satisfied, time dependent interactions
can be implemented. When only one static speckle pattern is implemented, the Hamiltonian describes disordered interactions.

\section{Experimental considerations}

%Now, we discuss some of the experimental considerations for the realisation of the 
%systems described above. We first compare our setup with the existing cQED experiments to assess the feasibility
%of the genuine supersolid transition we discussed above. Then we estimate the free space cooperativiy parameter $C$
%for our squeezed vacuum setup.

%
%Both in the case of a cavity or squeezed vacuum, the coupling between the enhanced vacuum fluctuations
%and atomic transitions has a considerable effect on the system dynamics only if the associated cooperativity
%satisfies $C \geq 1$. 

%
%\begin{figure}
%\includegraphics[width=0.5\textwidth]{Focusing}
%\caption{Focusing aparatus}
%\label{fig:focusing}
%\end{figure}
%In Appendix, we show that such an enhancement is indeed achievable 
%given that the collective enhancement factor due to collective coupling of the bosons to the squeezed 
%vacuum can be increased by loading more atoms in the squeezed vacuum beam (\textcolor{red}{but should 
%not the density play the important role here ? The coupling of each atom to the vacuum mode is lowered as 
%the mode volume is increased.})
%\begin{figure}
%\centering
%\includegraphics[width = 0.1\textwidth]{BECTransition} \quad\quad
%\includegraphics[width = 0.2\textwidth]{SpinTransition}
%\caption{Two level structures la la}
%\end{figure}

The figure of merit for the squeezed vacuum setup is the cooperativity
parameter $C$, which is the ratio of the decay rate into the squeezed vacuum 
modes to the decay rate into all other unsqueezed vacuum modes. Thus, it can be calculated by comparing the spectral 
densities of the squeezed and unsqueezed vacuum modes at the dipole transition frequency $\overline{\omega}$
\begin{align}
\nonumber C = \frac{J_{\mathrm{sq}}(\overline{\omega})}{J(\overline{\omega})}&=\frac{\sum'_{l}|\eta|^2e^{2s(\omega_l)}\delta(\overline{\omega}-\omega_l)}{\sum_{l,s} |\eta_{s}|^2\delta(\overline{\omega}-\omega_l)}\\
&\approx e^{2s(\overline{\omega})}\left(\frac{L^2}{A_{\mathrm{foc}}}\right)\frac{\sum'_{l}\delta(\overline{\omega}-\omega_l)}{\sum_{l,s} \delta(\overline{\omega}-\omega_l)}\equiv C_0 e^{2s(\overline{\omega})}
\label{eq:cooperativity}
\end{align}
where the primed sum goes only over the vacuum modes with non-zero squeezing, $L$ is the linear dimension of space and 
$A_{\mathrm{foc}}$ is the focal area transverse to the propagation direction of the squeezed vacuum mode.
The index $s$ denotes the polarization degree of freedom of the unsqueezed vacuum states.
For the second equality, we assumed that the bare dipole coupling stays approximately constant within 
 the squeezing bandwidth around the transition. In the last relation, we defined the cooperativity of
 unsqueezed modes $C_0$. Eq. ($\ref{eq:cooperativity}$) clearly shows that for the 
 dipole coupling to the squeezed vacuum modes to be significant, 
 the enhancement of fluctuations in the anti-squeezed quadrature should compensate for the reduction of interactions due to the finite 
 solid-angle subtended by the squeezed mode \cite{Tanji2011}. When the focal area of the unsqueezed vacuum beam
  $A_{\mathrm{foc}}$ is decreased down to $\overline{\lambda}^2 = \left(\frac{\overline{\omega}}{2\pi c}\right)^2$, the cooperativity is $C_0 = \frac{3}{8\pi}\approx 0.1$.
Thus, a cooperativity parameter of order 1 can be achieved even with a moderate anti-squeezing degree of 10 dB.
%
%
% In the cavity setting the cooperativity
%parameter is defined as $C_{\mathrm{cav}} \equiv \frac{\eta^2}{\gamma \kappa_c}$, where $\eta$ is the dipole 
%coupling strength at resonance, and $\gamma$ and $\kappa_c$ are the atomic and cavity decay rates, respectively. 
%For the squeezed vacuum setup, we use the free space cooperativity $C$, which is obtained by replacing
%$\kappa\rightarrow c/L$. In order for the squeezed vacuum to effectively modify the dynamics
%of the atomic system, cooperativity parameter should be at least of order 1.

It is also useful to calculate the ratio between the strength of the dispersive interactions to the decay rate.
In the weak coupling regime, the ratio is given by the dispersive cooperativity (see Appendix \ref{app:Greens})
\begin{align}
C_{\mathrm{disp}}(\omega_{\mathrm{dip}}) \approx \left(\frac{L^2}{\pi A_{\mathrm{foc}}}\right) \frac{\int d\omega \sum'_{l} \frac{2\omega e^{2s(\omega)}}{\omega_{\mathrm{dip}}^2-\omega^2}\delta(\omega-\omega_l)}{\sum_{l,s} \delta(\bar{\omega}-\omega_l)},
\end{align}
where we assumed that the transition frequencies of the dipoles that interact through the squeezed vacuum are identical 
and given by $\omega_\mathrm{dip}$.

\begin{figure}
\includegraphics{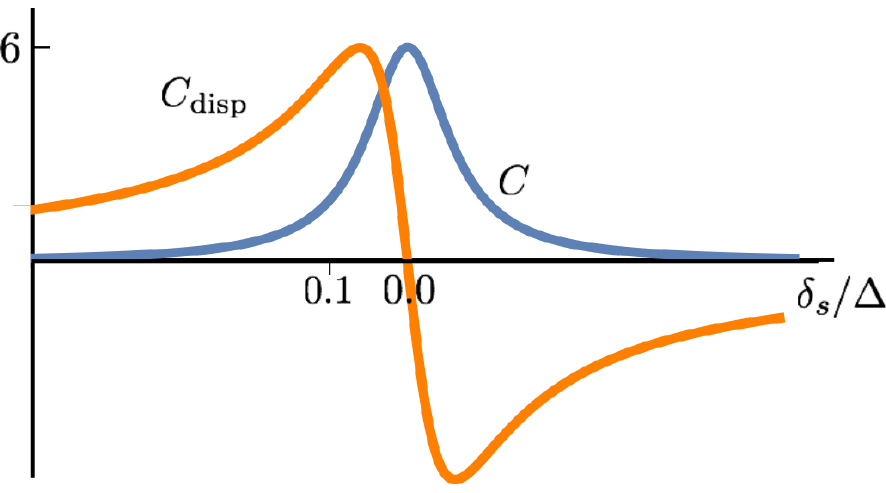}
\caption{``Color" The single atom cooperativity $C$ and the dispersive cooperativity $C_{\mathrm{disp}}$ as a function of the detuning $\delta_s$ between the squeezing and the dipole transition frequencies, normalized by the squeezing bandwidth $\Delta$. The squeezing parameter $s(\omega_{\mathrm{s}})$ is set to give 23 dB squeezing, and 50$\%$ losses is included. Note that the Lorentzian profile of the squeezing spectrum inevitably leads to a enhanced dissipation.}
\label{fig:enhancement}
\end{figure}

In an experimental situation, diffractive losses along the optical path of the squeezed vacuum beam as well as 
the Lorentzian profile
of the squeezing spectrum should be taken into account. While the diffractive losses
reduce the cooperativity parameter, the Lorentzian shape of the squeezing spectrum
unavoidably reduces the lifetime of the atomic transitions. In Fig. $\ref{fig:enhancement}$,
we show $C$ and $C_{\mathrm{disp}}$ as a function of $\delta_s = \omega_{s} - \omega_{\mathrm{dip}}$ 
for an experimental setup where $50\%$ of the squeezed vacuum is lost due to diffraction \cite{loudon1983quantum, Semmler2016}.
Moreover, the dipole transition frequency satisfies $\omega_{\mathrm{dip}}\ll \Delta$, such that the rotating-wave approximation 
is not valid for coupling to the squeezed vacuum modes (see Appendix $\ref{app:Greens}$).
Lastly, we note that by using additional squeezed vacuum beams, the spectral profile of 
squeezing can also be modified to further increase the dispersive interaction strength.

\section{Applications}
We now discuss some of the applications of the squeezed vacuum setup to realise tunable interactions
between spin and between density degrees of freedom in artificial quantum many-body systems. We also
show how the squeezed vacuum can be used to improve previous experimental proposals for realising exotic orders. 

For the implementation of tunable density-density interactions, we identify
\begin{align}
\hat{O}_{\mathrm{D}}^{\dagger}\equiv \frac{1}{2}(1-\sigma^{\mathrm{z}}) \equiv \psi_{g}^{\dagger}\psi_{g}, 
\label{eq:BEC}
\end{align}
where $\sigma^{\mathrm{z}}$ is a Pauli matrix which acts on the two dimensional  subspace spanned by
the ground ($g$) and excited ($e$) states of the atomic system. The resulting coupling has been realised in the recent 
experiments in cold atom systems \cite{Landig2016}. On the other hand, spin-spin 
interactions can be described by identifying
\begin{align}
\hat{O}_{\Lambda}^{\dagger}\equiv \sigma^{\mathrm{x}} \equiv \psi_{\uparrow}^{\dagger}\psi_{\downarrow}+\psi_{\downarrow}^{\dagger}\psi_{\uparrow},
\label{eq:spin}
\end{align}
where, $\sigma^{\mathrm{x}}$ is a Pauli operator which acts on two nearly degenerate hyperfine
levels. An implementation of this coupling in a 4-level atomic systems has been proposed by Ref. \cite{Dimer2007}.
We note that the above discussion can be extended to atomic systems with more internal degrees of freedom $d$. Here 
we restrict our attention to $d=1$ and $d=2$ for simplicity.

We emphasize the application of tunable density-density interactions to realize a continuous-space
supersolid transition in bosonic systems \cite{Prokofiev2012,Prokofiev2007}. Briefly, the continuous-space supersolid phase is characterised by the breaking of the continuous spatial translation symmetry and the $U(1)$ phase symmetry of bosonic fields. 

The continuous-space supersolid transition in a Bose Einstein condensate can be realised in the squeezed vacuum setup by 
using multiple squeezed vacuum beams in a traveling wave configuration. 
To this end, let us consider two traveling squeezed vacuum beams whose spatial profile on the two-dimensional
atomic system is given by
\begin{align}
\Xi_{l}^{\pm}(r) \propto e^{\pm i \mathbf{k}_0 \cdot r}.
\end{align}
Then, the squeezed vacuum induced interaction between an atom pair $\{i,j\}$ is 
\begin{align}
H_{int}^{ij} (\omega_l )\propto \cos{[\mathbf{k}_0 \cdot(\mathbf{r}_i-\mathbf{r}_j)]}\frac{1}{4}(1-\sigma_z^{i})(1-\sigma_z^{j}).
\end{align}
Thus, the density-density interactions have a characteristic momentum given by $\mathbf{k}_o$. If the interaction is attractive, this 
leads to the softening of the collective modes of the system with momentum $\mathbf{k}_o$. When the attractive interactions reach a 
critical strength the relevant soft modes become unstable, and a density modulation of the condensate with periodicity $\lambda^{-1}_{0} = |\mathbf{k}_0|/2\pi$ breaks the continuous translational symmetry of the condensate. In Appendix \ref{app:compcQED}, we show
that dipole coupling strengths comparable to a recent realisation of a lattice supersolid phase in a cQED setup \cite{Landig2016} 
can be achieved using the squeezed vacuum setup, by using a top-hat spatial profile for the squeezed vacuum beam.

Another application of the squeezed vacuum setup is the realisation of a spin-glass phase by tuning the exchange interactions 
between localised spins. The Hamiltonian that describes the dispersive interactions between any pair of qubits $(i,j)$ has the form
\begin{align}
H_{int}(r_i,r_j) &\propto \sum_{l} \Xi_l^*(r_i)  \Xi_l(r_j) \sigma_i^{\mathrm{x}} \sigma_{j}^{\mathrm{x}} 
\end{align} 
By using a static speckle pattern for the spatial profiles $\Xi_l$ of the squeezed vacuum beams, it is possible to 
realize a spin model with disordered interactions. As shown in Ref. \cite{Strack2011,Buchhold2013,Gopalakrishnan2011},
such a model exhibits a spin glass phase as well as ferromagnetic and paramagnetic phases.
We also note that a transverse exchange interaction can be implemented when the spins are coupled to the 
squeezed vacuum through a Jaynes-Cummings (JC) type coupling (see Appendix $\ref{app:QuadJC}$). It has been shown
that the transverse Heisenberg interaction is sufficient for an implementation of universal quantum 
computer \cite{Imamoglu1999}. For this implementation, as for the cycling rate for the implementation of finite-range interactions,
the gate ramp times should be slow enough to avoid transitions into states with non-zero number of excitations on
top of the squeezed vacuum state.
%
%\begin{figure}
%\includegraphics[width=0.5\textwidth]{Detection}
%\caption{Detection}
%\label{fig:Detection}
%\end{figure}

%
% As in the case of conserving interactions, the realisation of translationally invariant finite-range
% Lindbladian's is possible under the restriction that the system evolves under a time-averaged Lindbladian. 
% However, infinite range correlated decay processes can be interesting in their own right. For instance, super- and sub- 
% radiance phenomena have been observed in experiments with superconducting qubits placed on a transmission line
%  \cite{Loo2013}.

%
%\begin{align}
%C_{\mathrm{sq}} \equiv \frac{\tilde{g}}{\gamma_{\mathrm{dip}}},
%\label{eq:coop}
%\end{align}
%
%where $\tilde{g} = \sum_{l}\tilde{g}_l e^{2s(\omega_l)}$ is the effective coupling constant and $\gamma_{\mathrm{dip}}$ is the
%lifetime of the dipole transition whose operator is $\hat{O}$. We note that the lifetime of the dipole transition is also modified 
%by the presence of the squeezed vacuum. 

\section{Conclusion}
We propose squeezed vacuum as a valuable resource for engineering dispersive and dissipative 
interactions in artificial quantum many body systems. We show that experimentally demonstrated values of anti-squeezing
gives cooperativity parameters $C > 1$. The most important 
advantage of using squeezed vacuum engineering of the dipole interactions is that the spatial 
profiles of the electromagnetic vacuum modes can be modified dynamically in a traveling-wave geometry.
Thus, the vacuum modes can have an arbitrary profile in the plane perpendicular to the propagation direction.
Using this dynamical control, we showed how to implement arbitrary finite range interaction
potentials in a time-averaged fashion. 

We would like to emphasize that the idea of using the squeezed vacuum to engineer interactions between dipoles
is platform independent. Thus, it can be used to deterministically generate entanglement between any two atomic dipoles
whether they are implemented via quantum dots, NV centers \cite{Wrachtrup2006}, or real atoms, opening up new possibilities for  
hybrid systems. One exciting observation is that the entangling interactions can be implemented in macroscopic length 
scales as long as the distance between the dipoles is shorter than the coherence length of the squeezed vacuum, which
is determined by the squeezing bandwidth. Lastly, the enhanced fluctuations in the squeezed electromagnetic vacuum 
can be transferred to other vacuum fluctuations such as phonons in crystals or condensates. Thus, one can enhance the 
interactions between these phonons and other electronic degrees of freedom \cite{Cotlet2016,Knap2016}. 

\begin{acknowledgments} 
We acknowledge Sylvain Ravets, Ovidiu Cotlet, Emre Togan, Murad Tovmasyan, Evert van Nieuwenburg, 
Renate Landig, Lorenz Hurby, Nishant Dogra, Manuele Landini, Tobias Donner, Tilman Esslinger, Michael Gullans,
Matteo Maritelli, Jonathan Home, Aashish Clerk, Charles Edouard Bardyn for delightful discussions. 
\end{acknowledgments}

\appendix

\section{Squeezing spectrum}
\label{app:SqueezeSpec}
In this Appendix, we follow the standard derivation of the squeezing properties of the 
output field from an OPA \cite{gardiner2004quantum}.
\subsubsection{The equations of motion in frequency space}
Two mode squeezed light can be generated using parametric down conversion. 
Consider the Hamiltonian in the rotating frame of the pump frequency
\begin{align}
H_{PDC} = \hbar \omega_0(a^{\dagger}a + b^{\dagger}b) + i \hbar \left(\epsilon a^{\dagger}b^{\dagger}e^{-i\omega_p t} - \epsilon^* ab e^{+i\omega_p t}\right),
\end{align}
where $\omega_p$ is the OPA pump frequency and $\omega_0$ is the frequency of the OPA cavity modes.
We write down the Heisenberg-Langevin equations of motion for the cavity modes using $\frac{\partial A}{\partial t} = \frac{i}{\hbar} [H,A] $
\begin{align}
\dot{a}(t) &= -i\omega_0 a(t) + \epsilon b^{\dagger}(t)e^{-i\omega_p t} - \frac{\gamma_1}{2}a(t) - \sqrt{\gamma_1}a_{in}(t)\\
\dot{b}(t) &= -i\omega_0 b(t) + \epsilon a^{\dagger}(t)e^{-i\omega_p t} - \frac{\gamma_1}{2}b(t) - \sqrt{\gamma_1}b_{in}(t),
\end{align}
where $a_{in}$ and $b_{in}$ are the input noise operators for the corresponding cavity modes, and
we assumed that the cavity where the $\chi^{2}$ material is in has only one decay channel with rate $\gamma_1$.

We can solve this equation by going to the fourier domain
\begin{align}
a(t) = \frac{1}{\sqrt{2 \pi}}\int d\omega e^{-i\omega t} a(\omega)\\
[a(t)]^{\dagger} = a^{\dagger}(t) = \frac{1}{\sqrt{2 \pi}}\int d\omega e^{i\omega t} a^{\dagger}(\omega),
\end{align}
and the same for $b$. Specifically,
\begin{align}
\int dt \, b^{\dagger}(t)e^{i(\omega- \omega_p)t} = b^{\dagger}(\omega_p-\omega).
\end{align}
Then, taking the Fourier transform of the equation of motion, we get
\begin{align}
\nonumber  & i(\omega_p-\omega) a^{\dagger}(\omega_p-\omega) =\\
\nonumber  & +i\omega_0 a^{\dagger}(\omega_p-\omega) + \epsilon^* b(\omega) - \frac{\gamma_1}{2}a^{\dagger}(\omega_p-\omega) - \sqrt{\gamma_1}a_{in}^{\dagger}(\omega_p-\omega)\\
\nonumber &i(\omega_p-\omega) b^{\dagger}(\omega_p-\omega) =\\
\nonumber  & +i\omega_0 b^{\dagger}(\omega_p-\omega) + \epsilon^* a(\omega) - \frac{\gamma_1}{2}b^{\dagger}(\omega_p-\omega) - \sqrt{\gamma_1}b_{in}^{\dagger}(\omega_p-\omega)\\
\nonumber&-i\omega a(\omega) = -i\omega_0 a(\omega) + \epsilon b^{\dagger}(\omega_p-\omega) - \frac{\gamma_1}{2}a(\omega) - \sqrt{\gamma_1}a_{in}(\omega)\\
&-i\omega b(\omega) = -i\omega_0 b(\omega) + \epsilon a^{\dagger}(\omega_p-\omega) - \frac{\gamma_1}{2}b(\omega) - \sqrt{\gamma_1}b_{in}(\omega)\\
\end{align}
Now setting $\omega_p = 2\omega_0 \, \dot{=} \, 2 \omega_s$ we can write the following set  of linear equations as 
%\footnote{Going to the rotating frame of half the pump frequency, $a\rightarrow a e^{i\omega_p/2 t}$
%\begin{align}
%\dot{\mathbf{a}} = \left(A - \frac{\gamma_1}{2}\right) \mathbf{a} - \sqrt{\gamma_1}\mathbf{a}_{in},
%\end{align}}.
\begin{align}
\mathbf{a}(\omega) =\left( A - \left(\frac{\gamma
_1}{2} - i(\omega-\omega_s)\right)\right)^{-1} \sqrt{\gamma_1} \mathbf{a}_{in}(\omega),
\label{eq:RFvec}
\end{align}
where 
\begin{align}
\mathbf{a}(\omega) = \left(\begin{array}{c}a(\omega) \\ b(\omega)\\ a^{\dagger}(2\omega_s-\omega)\\ b^{\dagger}(2\omega_s-\omega)\end{array}\right)\quad \mathrm{and} 
\quad A = \left(\begin{array}{cccc} 0&0&0&\epsilon\\ 0&0& \epsilon& 0 \\ 0& \epsilon^{*}& 0 &0\\ \epsilon^{*} & 0 & 0 & 0 \end{array}\right).
\end{align}
The inverse of this matrix we need is
\begin{widetext}
\begin{align}
\nonumber &\left( A - \left(\frac{\gamma_1}{2} - i(\omega-\omega_s)\right)\right)^{-1} =  \frac{1}{(\frac{\gamma_1}{2} - i(\omega-\omega_s))^2 - |\epsilon|^2} \times \\ 
\nonumber &\left(\begin{array}{cccc} -(\frac{\gamma_1}{2} - i(\omega-\omega_s))&0&0&-\epsilon\\ 0&-(\frac{\gamma_1}{2} - i(\omega-\omega_s))&-\epsilon& 0 \\ 0& -\epsilon^*& -(\frac{\gamma_1}{2} - i(\omega-\omega_s)) &0\\ -\epsilon^* & 0 & 0 & -(\frac{\gamma_1}{2} - i(\omega-\omega_s)) \end{array}\right),
\end{align}
\end{widetext}
The intracavity fields can be written as
\begin{align}
a(\omega) &= \tilde{u}(\omega) a_{in} (\omega) - \tilde{v}(\omega) b_{in}^{\dagger}(2\omega_s-\omega)\\
b(\omega) &= \tilde{u}(\omega) b_{in} (\omega) - \tilde{v}(\omega) a_{in}^{\dagger}(2\omega_s-\omega)
\end{align}
where we defined
\begin{align}
\tilde{u}(\omega) = - \frac{(\frac{\gamma_1}{2} - i(\omega-\omega_s)) \sqrt{\gamma_1}}{(\frac{\gamma_1}{2} - i(\omega-\omega_s))^2 - |\epsilon|^2} \\
\tilde{v}(\omega)=\frac{\epsilon \sqrt{\gamma_1}}{(\frac{\gamma_1}{2} - i(\omega-\omega_s))^2 - |\epsilon|^2}.
\end{align}
Using this, we note the important symmetry of these so called coherence factors.
\begin{align}
\tilde{u}(\omega) &= \tilde{u}^*(2\omega_s- \omega)\\
\tilde{v}(\omega) &= e^{i2\theta_s} \tilde{v}^{*} (2\omega_s - \omega),
\label{eq:cohsym}
\end{align}
where we defined the squeezing angle $\theta_s$. The above relations reflect the energy conservation law for the squeezed excitation. That is, the excitations on top of the squeezed vacuum are a superposition of one photon excitation with frequency ($\omega_1$) and one photon hole with frequency ($\omega_2$) with the constraint  
\begin{align}
\omega_1 + \omega_2 = 2\omega_s = \omega_p.
\label{eq:encons}
\end{align}
The phase information in the imaginary part in the denominator simply reflects the $\pi$ shift over the 
resonance for $\tilde{u}(\omega)$, whereas $\tilde{v}(\omega)$ also includes the squeezing angle information.
For a single sided cavity, the squeezing angle is independent of $\omega$.

%\subsubsection{Broadband limit and the output field}
%We can investigate broadband squeezed light by going to the limit where $\gamma, \epsilon > \Delta_{range}$, where $\Delta_{range}$ is the bandwidth of the generated light\footnote{For the white noise limit we set $\Delta_{range}$ to be larger than the bandwidth of the 2DEG-exciton system.}.
%Moreover, $\gamma_1 > \epsilon$, because we would like to operate the parametric amplifier
%below oscillation threshold . Note that this corresponds to the situation in a coherently driven polariton condensate, where the condensate would have been unstable if it was not for the decay rate $\gamma_1$. In this limit, the $u(\omega) = u $ and $v(\omega) = v$ given by
%\begin{align}
%u(\omega) &= -\frac{\sqrt{\gamma_1} \left(\frac{\gamma_1}{2}\right)}{(\frac{\gamma_1}{2})^2 - |\epsilon|^2}\\
%v(\omega) &= \frac{\sqrt{\gamma_1}\epsilon}{(\frac{\gamma_1}{2})^2 - |\epsilon|^2}.
%\end{align}

\subsubsection{The output field of the OPA}
We are interested in the output field of the parametric down converter. This is obtained using the input-output relation, which reads
\begin{align}
\nonumber a_{out}(\omega)&\equiv \bar{a}(\omega)= a_{in}(\omega) + \sqrt{\gamma_1} a(\omega) \\
&=u(\omega) a_{in} (\omega) - v(\omega)  b_{in}^{\dagger}(2\omega_s-\omega) .  
\label{eq:OutFields}
\end{align}
where $u$, $v$ are
\begin{align}
\nonumber u(\omega)&= 1+ \sqrt{\gamma_1}\tilde{u}(\omega)=1 - \frac{(\frac{\gamma_1}{2} - i(\omega-\omega_s)) \gamma_1}{(\frac{\gamma_1}{2} - i(\omega-\omega_s))^2 - |\epsilon|^2}  \\
&=- \frac{(\frac{\gamma_1}{2})^2 + (\omega- \omega_s)^2+ |\epsilon|^2}{(\frac{\gamma_1}{2}-i(\omega-\omega_s))^2 -|\epsilon|^2}\\
v(\omega)&=\sqrt{\gamma_1}\tilde{v}(\omega) = \frac{\gamma_1 \epsilon}{\left(\gamma_1/2-i(\omega-\omega_s)\right)^2-|\epsilon|^2}
\label{eq:uvOut}
\end{align}
Note that the symmetries in Eq. ($\ref{eq:cohsym}$) are preserved, and now $$|u(\omega)|^2-|v(\omega)|^2=1.$$
As a result, we simply obtain a squeezed vacuum in terms of the input noise fields. Also note that
the large squeezing limit happens when $\epsilon$ approaches $\gamma/2$ from below. In this limit
$|u(\omega)|\approx |v(\omega)|$. 
%\begin{figure}
%\centering
%\includegraphics[width=0.5 \textwidth]{uvmag}
%\caption{The squeezing degree $e^{2s} = 10^{XdB/10}$ as a function of frequency for $\gamma = 0.6 \omega_s$ and $\epsilon = 0.2 \omega_s$. The condensate phase is set to $\theta=0$. But note that this is a very crude
%model, so in the following, we will simply take numbers from the experiments to get the bandwidth and the squeezing degree.}
%\label{fig:uvmag1}
%\end{figure}
Also note that the normalisation ensures that the inverse transform reads
\begin{align}
a_{in} (\omega) =u(\omega) \bar{a} (\omega) + v(\omega)  \bar{b}^{\dagger}(2\omega_s-\omega),
\end{align}
In order to take into 
account the diffractive losses we need to use the following expression for the output field \cite{loudon1983quantum}.
\begin{align}
\bar{a}= \sqrt{\epsilon_{\mathrm{diff}}} \left(a_{\mathrm{in}}(\omega) + \sqrt{\gamma_1}a(\omega)\right) + \sqrt{1-\epsilon_{\mathrm{diff}}} c(\omega),
\label{eq:sqout}
\end{align}
where $c(\omega)$ represent the noise that is injected into the squeezed vacuum due to losses, and 
$\epsilon_{\mathrm{diff}}$ is the power loss due to diffraction.
It has the noise properties of unsqueezed vacuum and obeys
\begin{align}
[a(\omega),c(\omega')] = [a_{\mathrm{in}}(\omega),c(\omega')] = 0.
\end{align}
The type of noise injected into the squeezed vacuum due to losses can be specified by 
specifying the correlation functions for $c(\omega)$.
%\begin{align}
% a_{out}(\omega)&= a_{in}(\omega) + \sqrt{\epsilon_{\mathrm{diff}}}\sqrt{\gamma_1} a(\omega)\\
% & \textcolor{red}{+ \mathrm{Should I have to multiply the whole thing by } \sqrt{\epsilon} ?}  ,
%\end{align}
This completes our discussion of the generation of squeezed vacuum using an OPA.
%\section{Squeezing transformation on the interaction Hamiltonian}
%For the condensate model, we start from the sum
%\begin{align}
%\nonumber &\sum_{\sigma=\pm} \eta_l \left( \Xi^{\sigma}_{l}(r) a^{\sigma}_{in}(\omega_l)  + (\Xi_{l}^{-\sigma})^*(r) (a^{-\sigma})^{\dagger}_{in}(\omega_l) \right),
%\end{align}
%and substitute
%\begin{align}
%\Xi^{\sigma}_{l}(r) a^{\sigma}_{in}(\omega_l) = u_{l} \Xi_{l}^{\sigma}(r) a^{\sigma}(\omega_l)  + v_l (\Xi_{l}^{-\sigma})^*(r) (a^{-\sigma})^{\dagger}(\omega_l),
%\end{align}
%and  
%\begin{align}
%\nonumber (\Xi_{l}^{-\sigma})^*(r) (a^{-\sigma})^{\dagger}_{in}&(\omega_l) =\\
% & u_{l}^* (\Xi_{l}^{-\sigma})^*(r) (a^{-\sigma})^{\dagger}(\omega_l)  + v_l^* \Xi_{l}^{\sigma}(r) a^{\sigma}(\omega_l)
%\end{align}
%collecting terms we obtain
%\begin{align}
%\sum_{\sigma=\pm} \eta_l \left( \kappa_l \Xi^{\sigma}_{l}(r) a^{\sigma}(\omega_l)  +  \kappa^*_l (\Xi_{l}^{-\sigma})^*(r) (a^{-\sigma})^{\dagger}(\omega_l) \right),
%\end{align}
%where $$\kappa_l \equiv (u_l + v_l^*)$$
\section{Coupling to squeezed vacuum source}
\label{app:Cascade}
In this Appendix, we derive the interaction Hamiltonian in Eq. $\ref{eq:BosHam}$ between the squeezed vacuum generated by an OPA 
and an atom which overlaps with the output of the OPA. To this end, we follow the work of Gardiner \cite{Gardiner1993}
and Carmichael \cite{Carmichael1993} on cascaded quantum systems.

The Hamiltonian that we are interested in is
\begin{align}
\nonumber H/\hbar & = H_{sys}/\hbar + \int d l \omega_l b_{l}^{\dagger}b_l + \int dl \kappa_c (l)(a b_l^{\dagger} + h.c.) \\
&+  \int d_l \kappa_a(l) (\sigma^{-}b_l^{\dagger}e^{-i\omega_l \tau} + h.c.),
\end{align}
where $l$ labels the vacuum modes that serve as the common bath modes for the OPA cavity and the atom, whose creation(annihilation)
operators are given by $a^{\dagger}(a)$ and $\sigma^{+}(\sigma^{-})$, respectively. The coupling constants $\kappa_a$ and $\kappa_c$ depend on $l$ in general.
The extra phase factor, $e^{-i\omega_l \tau}$ on the atom-field coupling is due to the distance between the source and the atom. 
For simplicity, we have not considered the quadrature coupling in this section, but the generalization is immediate.

Assuming now that the coupling of the cavity to the electromagnetic field modes is Markovian [i.e., $\kappa_c(l) \equiv \sqrt{\gamma_1}$], 
the equation of motion for the electromagnetic field has the solution
\begin{align}
\nonumber & b( l , t)  = e^{-i \omega_l (t-t_0)} b(l,t_0 ) \\
&+ \int_{t_0}^{t} dt' e^{-i\omega_l (t-t')}\left[ \sqrt{\gamma_1}a(t') + \tilde{\kappa}_a(l) \sigma^{-}(t')\right],
\end{align}
where $\tilde{\kappa}_a(l) \equiv \kappa_a(l)e^{-i\omega_l \tau}$.
On the other hand, the equation of motion of the atomic degree of freedom is given by 
\begin{align}
\dot{\sigma}^{-}(t) =  - i \omega_{\mathrm{at}} \sigma^{-}(t) - i \int d l \tilde{\kappa}_a(l) b(l,t),
\end{align} 
with the atomic transition frequency $\omega_{\mathrm{at}}$. Plugging in the solution for the electromagnetic field, one obtains
\begin{align}
\nonumber &\dot{\sigma}^{-}(t) =  - i \omega_{at} \sigma^{-}(t) - i \int d l \tilde{\kappa}_a(l) \bigg\{e^{-i \omega_l (t-t_0)} b(l,t_0 )\\
& + \int_{t_0}^{t} d t' e^{-i\omega_l (t-t')}\left[ \sqrt{\gamma_1}a(t') + \tilde{\kappa}_a(l) \sigma^{-}(t')\right]\bigg\},
\end{align}
whose fourier transform is
\begin{align}
\nonumber & i(\omega -\omega_{\mathrm{at}})\sigma^{-}(\omega) = i \int_{t_0}^{\infty}  dt \bigg( \int d l \tilde{\kappa}_a(l) \bigg\{e^{i\omega t}e^{-i \omega_l (t-t_0)} b(l,t_0 ) \\
 & + \int_{t_0}^{t} dt' e^{i\omega t} e^{-i\omega_l (t-t')}\left[ \sqrt{\gamma_1}a(t') + \tilde{\kappa}_a(l) \sigma^{-}(t')\right]\bigg\}\bigg),
\end{align}
The first term in the right hand side can be related to the $l$ components of the noise operator,
\begin{align}
\int dl \tilde{\kappa}_a(l)e^{i\omega_l t_0} b(l,t_0) \delta(\omega-\omega_l) \equiv \int dl \tilde{\kappa}_a(l) a_{in}(l) \delta(\omega-\omega_l) ,
\end{align}
which couples only at frequency $\omega_l$.
To evaluate the second term, we need a couple of assumptions. Re-arranging and setting $t_0 = 0$, we have
\begin{align}
\int d l \sqrt{\gamma_1} \tilde{\kappa}_a(l) \int_{0}^{\infty} dt e^{i(\omega-\omega_l) t} \int_{0}^{t} dt'  e^{i\omega_l t'} a(t') ,
\end{align}
the rightmost integral is evaluated by plugging in the fourier expansion of $a(t') = \int_{-\infty}^{\infty} d\omega' \exp{(-i\omega' t' )} a(\omega') $
\begin{align}
\nonumber \int_{0}^{t} dt'  e^{i\omega_l t'} a(t') &=\int_{-\infty}^{\infty} d\omega' a(\omega')  \int_{0}^{t} dt'  e^{i(\omega_l-\omega')t'} \\
& = \int_{-\infty}^{\infty} d\omega' a(\omega')  \int_{-\infty}^{\infty} dt'  e^{i(\omega_l-\omega')t'} \mathrm{Sq}(0,t),
\end{align}
where $\mathrm{Sq}(0,t)$ is a square function from $0$ to $t$. Note that $t$ gives us the time scale where we would like to describe the 
dynamics of the dipole in question. In steady state, we can take $t$ as far from $0$ as we want. Correspondingly, in steady state, we obtain
\begin{align}
\nonumber&\int_{-\infty}^{\infty} d\omega' a(\omega')  \int_{-\infty}^{\infty} dt'  e^{i(\omega_l-\omega')t'} \mathrm{Sq}(0,t) \\
 & \approx  \int_{-\infty}^{\infty} d\omega' a(\omega') \delta(\omega_l-\omega') = a(\omega_l) ,
\end{align}
where the $\delta$ function actually has the width of $2\pi/t$. The second integral over $t$ gives an additional factor of $\delta(\omega_l~-~\omega)$. Thus,
the electromagnetic vacuum couples to the dynamics in an energy-conserving way.
As a result, we obtain,
\begin{align}
\nonumber -i\omega \sigma^{-}(\omega) &= - i \omega_{at} \sigma^{-}(\omega) -i \int dl \tilde{\kappa}_a(l) \left[ a_{in}(l) +\sqrt{\gamma_1} a(\omega_l)  \right]\\
& = - i \omega_{at} \sigma^{-}(\omega) -i \int dl \tilde{\kappa}_a(l) a_{\mathrm{out}}(\omega_l),
\end{align}
where $a_{\mathrm{out}}$ is defined as in Eq. ($\ref{eq:OutFields}$), 
and we have included in the renormalized transition frequency
$\tilde{\omega}_{at}~=~\omega_{at} + \Delta_{\mathrm{Lamb}} - i \gamma_{at}$,
the Lamb shift $\Delta_{\mathrm{Lamb}}$ and 
spontaneous decay rate $\gamma_{at}$
of the atomic transition due to its coupling to the \textit{free} electromagnetic modes.
The interaction Hamiltonian that produces this equation of motion is 
\begin{align}
H_I^{eff}/\hbar = \int dl \tilde{\kappa}_a(l) a_{\mathrm{out}}(\omega_l)\sigma^{+} + h.c. \, .
\end{align}
Notice the above equation is what we have used in the main text when the diffractive losses
replace $a_{out}\rightarrow \bar{a}$ as in Eq. ($\ref{eq:sqout}$).

In the cascaded system the chirality of the interactions, that is , the fact that the OPA field can influence the atom
without being influenced by the atom, can be seen by restricting the $l$ summation over the modes to ones whose 
wavevectors satisfy $k_l\cdot{(r_{\sigma}-r_{a})} > 0$, where $r_{\sigma}-r_a$ is the displacement vector from the 
cavity and the atom. For simplicity, let us consider the one dimensional situation where we consider the bath modes $b_l$
which are plane waves whose wavenumber $k_l$ is aligned with  $(r_{\sigma}~-~r_a) = c\tau \hat{x}$ with $c\tau > 0 $. Then, defining the spatial fourier transform 
of the common bath modes $b(l,t)$ as
\begin{align}
b(x , t) = \int d l b(l , t) e^{i \omega_l / c x},
\end{align}
where $x$ is measured with respect to the cavity, we find that  the contribution of the OPA to  $b(r , t)$  is 
\begin{align}
\nonumber b(x,t) \bigg | _{a} &=\int_{t_0}^t dt' \Theta(t-t') \int d l  e^{i \omega_l  (x/c - t + t' )}\sqrt{\gamma_1}a(t') \\
&=\Theta(x/c) \sqrt{\gamma_1} a(t - x/c),
\end{align}
where the contribution with $x \rightarrow -x$ is absent because of the restriction $k_l > 0$.
Thus, the influence of the cavity is only to the systems that are to its right. 

Similarly, the influence of the 
atom also propagates to the right since the contribution to the bath modes is 
\begin{align}
b(x,t) \bigg | _{\sigma} =   \int_{t_0}^{t}& dt' \Theta(t-t') \sigma^{-}(t') \tilde{\kappa}_a(t-t'-x/c + \tau),
\end{align}
where $\tilde{\kappa}_a(x)\equiv \int dl e^{i\omega_l x}\kappa_a(l)$.
We note that the contribution of the off-resonant atom is non-Markovian.
To recover the Markovian expression, we replace $\kappa_a(l)$ by $\sqrt{\gamma_a}$, and
 $\tilde{\kappa}_a (t-t'-x/c + \tau)$ by $\delta(t-t'-x/c + \tau)$, to obtain familiar contribution \cite{loudon1983quantum}
\begin{align}
b(x,t) \bigg | _{\sigma} = \Theta(x/c - \tau) \sqrt{\gamma_a} \sigma^{-} (t - x/c + \tau).
\end{align}
For a more non-trivial case of the coupling function $\kappa_a(l)$, let us consider 
$\kappa_a(l) \equiv \frac{\gamma}{\gamma^2 + (\omega_s - \omega_l)^2}$, where $\omega_s$, and $\gamma$
can be thought of as the carrier frequency and the bandwidth of the squeezed vacuum, respectively. Then the contribution of the 
atom to the common bath mode can be calculated using the residue theorem.
\begin{align}
\nonumber b(x,t) \bigg | _{\sigma} &\propto \\
\nonumber &\int_{t_0}^{t} dt' \bigg\{ \Theta(t-t')  \sigma^{-}(t') e^{-(\gamma - i \omega_s)|t-t'-x/c+\tau| } \bigg\}.
\end{align}
Now let us look at the field distribution around the position of the atom. With $\delta x \equiv x - c \tau  $,
$\delta t  \equiv t - t'$, and a shift of the time integral to the integration variable $\bar{t} \equiv \delta t - \delta x $, we get
($c\equiv1$)
\begin{align}
b(x,t) \bigg | _{\sigma} &\propto \int^{t-t_0 - \delta x}_{-\delta x} d\bar{t}  \Theta( \delta x +\bar{t})  \sigma^{-}(t - \bar{t}- \delta x) e^{-(\gamma - i \omega_s)|\bar{t}| },
\end{align}
Unlike the Markovian case, the atomic contribution to the bath mode at time $t$ and $x$ 
is non-vanishing for $ |\delta x|~>~0 $ although the 
bath modes that couple to the atom have a definite chirality. However, as expected, the influence of the atom extends more to the right
($\delta x >0$) than to the left ($\delta x < 0$), due to the exponential factor in the integrand.

Lastly, we note that in the main text, we incorporated the free evolution of the output field of the OPA by 
$\hbar\int dl \omega_l a^{\dagger}_{\mathrm{out}}(l)a_{\mathrm{out}}(l)$, 
which holds thanks to the energy conservation in the steady state which enabled us to employ delta functions above.

\section{Green's functions}
\label{app:Greens}
Here we give a brief introduction to the Green's function methods \cite{cohen1992atom} to describe the effective 
dynamics of the atomic system placed within the squeezed vacuum. We project the many-body Schroedinger
equation onto the manifold of atomic and the squeezed vacuum states that are of interest. Then we derive the effective Hamiltonian
for these atomic states in the pole approximation. 
%We also elaborate on the conditions for the projection onto the 
%squeezed vacuum to be valid. The validity of the projection determines the validity of neglecting the AC Stark shift
%terms in Eq. (\ref{eq:DDInt}). 
The effective interactions between dipoles placed in the squeezing vacuum can be described by the 
so called level shift operator projected onto the subspace  $\{\alpha\}$ of the many body atomic states
tensored with the squeezed vacuum state in the suitable rotating frame. We note that the 
derivation of the effective Hamiltonian using the Green's function methods does not rely on a rotating-wave
approximation.

To this end, we define the projection operator onto $\{\alpha\}$
\begin{align}
P_{\{\alpha\}} &= \sum_{ i \in \{\alpha\}} |i \rangle\langle i |\\
Q_{\{\alpha\}} &= \mathbf{I} - P_{\{\alpha\}} = \sum_{i\not\in \{\alpha\} }|i \rangle\langle i |.
\label{eq:projector}
\end{align}
The projector operators obey $P^2_{ \{ \alpha \}}=P_{\{\alpha\}}$, $Q^2_{\{\alpha\}}=Q_{\{\alpha\}}$, and
$Q_{\{\alpha\}}P_{\{\alpha\}} =0 $. Moreover, $P_{\{\alpha\}} H_{0} Q_{\{\alpha\}}  = 0 $. 
 
Now, the Green's function operator is defined as 
\begin{align}
(z-H)G(z) =(z-H_0 - V)G(z) = \mathbf{I},
\end{align}
where $z \in \mathbf{C}$. 
Multiplying the above equation by $P_{\{\alpha\}}$ from the right, we obtain
\begin{align}
(z-H_0 - V)(P^2_{\{\alpha\}}+ Q^2_{\{\alpha\}})G(z)P_{\{\alpha\}} = P_{\{\alpha\}}.
\label{eq:projSch}
\end{align}
Then multiplying the above equation on the left with $Q_{\{\alpha\}}$ and using the properties of the projector operators
that are mentioned above, we obtain
\begin{align}
Q_{\{\alpha\}} G(z) P_{\{\alpha\}} = \frac{Q_{\{\alpha\}} VP_{\{\alpha\}}}{Q_{\{\alpha\}}(z-H)Q_{\{\alpha\}}} (P_{\{\alpha\}} G(z)P_{\{\alpha\}}).
\label{eq:aboveEq}
\end{align}
On the other hand, multiplying Eq. ($\ref{eq:projSch}$) on the left with $P_{\{\alpha\}}$ and substituting in Eq.~\ref{eq:aboveEq} , 
we find that the projected Greens function operator satisfies the following equation
\begin{align}
P_{\{\alpha\}}\left[ z-H-V \frac{Q_{\{\alpha\}}}{Q_{\{\alpha\}} (z-H)Q_{\{\alpha\}}}V\right]P_{\{\alpha\}}G(z)P_{\{\alpha\}}=P_{\{\alpha\}}.
\end{align}
Thus, the projected Green's function operator has the form
\begin{align}
G_{\{\alpha\}}(z) \equiv P_{\{\alpha\}} G(z)P_{\{\alpha\}} = \frac{P_{\{\alpha\}}}{P_{\{\alpha\}}(z-H_0-R(z)) P_{\{\alpha\}}},
\end{align}
where 
\begin{align}
R(z) = V + V\frac{Q_{\{\alpha\}}}{Q_{\{\alpha\}}(z-H)Q_{\{\alpha\}}}V,
\end{align}
is called the level-shift, or the self-energy operator.
The level-shift operator can be expanded in powers of the interaction Hamiltonian $V$
\begin{align}
\nonumber R(z)& = V + V\frac{Q_{\{\alpha\}}}{z-Q_{\{\alpha\}}H_0Q_{\{\alpha\}}}V \\
&+ V\frac{Q_{\{\alpha\}}}{z-Q_{\{\alpha\}}H_0Q_{\{\alpha\}}}V\frac{Q_{\{\alpha\}}}{z-Q_{\{\alpha\}}H_0Q_{\{\alpha\}}}V + \cdots.
\end{align}
So far, we have kept the discussion very general. However, the level - shift operator is greatly 
simplified when the interaction Hamiltonian only couples the bath and
atomic degrees of freedom such that the terms proportional to $Q_{\{\alpha\}}VQ_{\{\alpha\}}$ vanish. This 
results in
\begin{align}
R(z) = V + V\frac{Q_{\{\alpha\}}}{z-Q_{\{\alpha\}}H_0Q_{\{\alpha\}}}V.
\end{align}
The real and the imaginary parts of the shift operator are found using the Sokhotski-Plemelj identity
\begin{align}
\frac{1}{x\pm i 0^{+}} = \mathcal{P}\frac{1}{x} \mp i \pi \delta(x),
\end{align}
where $\mathcal{P}$ denotes the principle value.
Then, near the real number line
the Hermitian and the anti-Hermitian parts of the shift operator are given by
\begin{align}
\nonumber R(E\pm i0^+) &= V + \mathcal{P}V\frac{Q_{\{\alpha\}}}{Q_{\{\alpha\}}(E \pm i0^+ -H_0)Q_{\{\alpha\}}}V \\
\nonumber&\mp i \pi VQ_{\{\alpha\}}\delta(E-Q_{\{\alpha\}}H_0Q_{\{\alpha\}})Q_{\{\alpha\}}V\\
& = V + \Sigma(E\pm i 0^{+}) \mp i \hbar \Gamma(E)/2 .
\label{eq:shiftop}
\end{align}
It is clear that the shift operator changes the energy levels and introduces decay rates inside the 
projected state space $\{\alpha\}$ due to the interactions with the states outside of $\{\alpha\}$. 
If the Hermitian part of the shift operator is small compared
to the bare energies $\{E_{\alpha}\}$ of the states in $\{\alpha\}$, one can apply the so called pole approximation
to arrive at the approximation of the Green's function 
\begin{widetext}
\begin{align}
 G_{\{\alpha\}}(E\pm i0^+)& \approx \frac{ P_{\{\alpha\}} }{P_{\{\alpha\}}(E - H_0- V - \Sigma(\{E_{\alpha}\}) \mp i( \frac{\hbar}{2}\Gamma(\{E_{\alpha}\}) + 0^+))P_{\{\alpha\}}} \equiv  \frac{ P_{\{\alpha\}} }{P_{\{\alpha\}}(E - H_{\{\alpha\}} +i 0^+)P_{\{\alpha\}}},
\end{align}
\end{widetext}
where we defined the projected Hamiltonian
\begin{align}
H_{\{\alpha\}}(\{E_{\alpha}\}) = H_0 + V+ \Sigma(\{E_{\alpha}\}) + i \frac{\hbar}{2}\Gamma(\{E_{\alpha}\}).
\label{eq:EffHam}
\end{align}
In the above equations, expressions $\Sigma(\{E_{\alpha}\})$ and $\Gamma(\{E_{\alpha}\})$ take 
$E_{\alpha}$ as their arguments depending on the specific state $\alpha$ that $P_{\{\alpha\}}$ projects 
the interaction term onto.

The matrix elements of the effective Hamiltonian can now be calculated. As mentioned in the main text, we 
are interested in the interactions that are within the zero photon manifold. For instance the matrix element 
associated to the exchange of an excitation between two dipoles with transition frequency $\omega_{\mathrm{dip}}$ is given by 
\begin{align}
\nonumber &\langle g,e; \{0_l\} | \Sigma(\omega_{\mathrm{dip}}) |e,g,; \{0_l\} \rangle\\
\nonumber &= \langle g,e; \{0_l\} | V \frac{1}{\omega_{dip} - H_0} V |e,g,; \{0_l\} \rangle\\
\nonumber&=\sum_{l} g_l (r_1)g_{l}(r_2)\left( \frac{1}{\omega_{\mathrm{dip}} - \omega_{l}} + \frac{1}{\omega_{\mathrm{dip}} - (\omega_{l} + 2\omega_{\mathrm{dip}})}\right)\\
&= \sum_{l}  g_l (r_1)g_{l}(r_2) \frac{2\omega_{l}}{\omega_{\mathrm{dip}}^2 - \omega_l^2}, 
\end{align}
where the intermediate two-qubit states in the one photon manifold
have energies  $H_0 |g,g; 1_l\rangle = \omega_{l}$ and $H_0 |e,e;1_l\rangle = (\omega_{l} + 2\omega_{\mathrm{dip}})|e,e;1_l\rangle$.
We also assumed that dipolar coupling strength $g_l$ at positions $r_1$ and $r_2$ are real for simplicity. We took the interaction potential 
between the bath and the two dipoles 
to be of the form $V~=~\sum_{l} \left( g_l(r_1)\sigma_1^{+}X_l + g_{l}(r_2)\sigma_2^+X_l \right) $.

Using the above procedure, it is also easy to show that a thermal state, or a Fock state of photons does not increase the effective coupling 
strength. To this end, consider an initial state where each electromagnetic mode $l$ is in a Fock state with $n_l$ photons, the intermediate states that need to
 be taken into account are: $|e,e; \{n_l;n_{l'}+1\}\rangle$, $|e,e; \{n_l ;n_{l'}-1 \}\rangle$, $|g,g; \{n_l ;n_{l'}+1\}\rangle$, $|g;g;\{ n_l;n_{l'}-1\}\rangle$. Here we used the 
following notation 
 \begin{align}
 |g;g;\{ n_l;n_{l'}-1\}\rangle = |g\rangle \otimes |g\rangle \otimes |n_1,n_2,\cdots,n_{l'}-1,\cdots \rangle.
\end{align}
 These states have the bare energies
$H_0 |g,e; \{n_l\}\rangle  \equiv H_0 |e,g; \{n_l\}\rangle  = \omega_{\mathrm{dip}} +\sum_l n_l \omega_l $. Thus,
the matrix element of $\Sigma(\omega_{\mathrm{dip}})$ becomes
\begin{align}
\nonumber &\langle g,e; \{n_l \}| \Sigma\left(\omega_{\mathrm{dip}}+ \sum_{l}\omega_{l}n_l \right) |e,g,; \{n_l\} \rangle\\
\nonumber& = \sum_{l}(n_l+1) g_l^2 \left[ \frac{1}{\omega_{\mathrm{dip}} - \omega_l } +\frac{1}{\omega_{\mathrm{dip}} -  \omega_l -2\omega_{\mathrm{dip}}}\right]\\
\nonumber & -( n_l) g_l^2 \left[ \frac{1}{\omega_{\mathrm{dip}} - \omega_l } +\frac{1}{\omega_{\mathrm{dip}} - \omega_l -2\omega_{\mathrm{dip}} }\right]\\
\nonumber & = \sum_{l} g_l^2 \frac{2\omega_{l}}{\omega_{\mathrm{dip}}^2 - \omega_l^2},
\end{align}
and we observe that the strength of the effective interactions does not depend on $\{n_{l}\}$.
Yet, the strength of the interaction mediated by squeezed Fock or thermal states is still enhanced 
since then we should replace $g_{l} \rightarrow g_{l}e^{s(\omega_l)}$. 
We emphasize that the above procedure does not assume rotating-wave or Markov approximations.

On the other hand, the decay rate is concerned with a transition between two states with different photon numbers. If we consider 
the quantity $|\langle e,0|G(E+i0^+)|e,0\rangle|^2$, and transform the expression back into the time domain, we obtain the decay rate of the population to be
\begin{align}
\Gamma(\omega_{\mathrm{dip}}) = \frac{2\pi}{\hbar}  \sum_{l}g_l^2 \delta(\omega_l - \omega_{\mathrm{dip}}).
\end{align}

The cooperativity $C$ and dispersive cooperativity $C_{\mathrm{disp}}$ were calculated by comparing the self-energies induced by the 
squeezed vacuum focused on the atoms to the imaginary part of the self energy induced by the coupling of the atom to the unsqueezed 
vacuum modes.

Lastly, we would like to note that although we have assumed the pole approximation here, the problem of a 
single emitter coupled to a squeezed vacuum quadrature is indeed exactly solvable without the rotating-wave or Markov 
approximations \cite{cohen1992atom}. Here our focus has been on the potential of squeezed vacuum states, and not the solution to a specific problem.

To summarise, we briefly introduced the Green's function framework to describe the effective 
dynamics of the atomic degrees of freedom in the presence of a squeezed vacuum.

\section{Quadrature vs. Jaynes-Cummings Hamiltonians}
\label{app:QuadJC}
In this Appendix, we extend our analysis of the quadrature coupling between the dipole transitions
and the vacuum modes to a Jaynes-Cummings type interaction where the contribution of anti-resonant terms are ignored.
We show that in the simplest case of a single squeezed mode, the enhancement factor is reduced by half, as the dipole transition
couples to both the squeezed and anti-squeezed vacuum fluctuations. 

For the Dicke model, the coupling of the atomic dipole is to the quadrature of the 
vacuum mode. Then the action of squeezing transformation on the quadrature operator 
is 
\begin{align}
\nonumber \bar{X}(r) &= \sum_{l} U^{\dagger}_{l}X_{l}(r)U_{l}  =\sum_{l} U^{\dagger}_{l}\frac{\Xi_l(r)}{2}\left(a_{l} + a_{-l}^{\dagger}\right)U_{l}\\
\nonumber& = \frac{\Xi_l(r)}{2} \sum_{l} \left( u_{l}\bar{a}_{l} + v_{l}\bar{a}_{-l}^{\dagger} + u^{*}_{-l}\bar{a}^{\dagger}_{-l}+v_{-l}^{*}\bar{a}_{l} \right)\\
& = \frac{\Xi_l(r)}{2} \sum_{l} \left((v_{-l}^{*} + u_{l})\bar{a}_{l} + (v_{l}+u^{*}_{-l})\bar{a}_{-l}^{\dagger} \right),
\label{eq:quad}
\end{align}
where  the pairs of labels $\left\{ l, -l \right\}$ correspond to frequency pairs across the 
squeezing frequency $\omega_s$ and we assumed that the spatial wave functions of the modes
symmetric across the squeezing frequency are the same (i.e., $\Xi_l(r)=\Xi_{-l}(r)$ and $\bar{a}(\omega)\equiv \bar{b}(\omega)$).
For simplicity, we assume the $\theta_s = 0 $, then using the relations in Eq. ($\ref{eq:cohsym}$)
% Writing the above coefficients in the polar form 
%$$u_{l} + v^*_{-l} = |u_{l} +v^*_{-l}| e^{i\theta}, $$
we arrive at the transformation rule
\begin{align}
\sum_{l} \bar{X}_l = \frac{1}{2} \sum_{l} (u_{l} + v_{l})(\bar{a}_{l} + \bar{a}^{\dagger}_{-l}).
\end{align}

We have so far considered the action of the squeezing transformation on the quadrature defined in 
Eq. ($\ref{eq:quad}$). However, in the rotating frame, a Jaynes-Cummings(JC) type coupling between the 
dipoles and the squeezed vacuum modes is also relevant \cite{Gardiner1986}. The JC interaction Hamiltonian is given by 
\begin{align}
H_{\mathrm{JC}}/\hbar = \sum_{l}\eta_l \left( \Xi_{l}(r) \hat{O}^{\dagger}(r) a_l + h.c.\right),
\end{align} 
transforms under the squeezing transformation as
\begin{align}
\nonumber \sum_{l} & \eta_l \bigg( \hat{O}^{\dagger}(r) (u_{l} \Xi_{l}(r) \bar{a}_l  + v_{-l} \Xi_{l}^*(r) \bar{a}^{\dagger}_{-l})  \\ 
\nonumber & +\hat{O}(r)(u_{l}^* \Xi_{l}^*(r) \bar{a}^{\dagger}_l  + v_{-l}^* \Xi_{l}(r) \bar{a}_{-l}) \bigg)\\
\nonumber &=\sum_{l}  \eta_l  \sqrt{|u_{l}|^2+|v_{l}|^2} \bigg(   \bar{a}_l \tau_{l}^{\dagger}(r) + \bar{a}^{\dagger}_{-l} \tau_{l}(r)  \bigg),
\end{align}
where we again assumed $\Xi_l(r) = \Xi_{-l}(r)$ and defined
\begin{align}
\nonumber\tau_{l}^{\dagger}(r) &=\Xi_l(r)\left(\frac{u_{l}\hat{O}^{\dagger}(r) }{\sqrt{|u_{l}|^2+|v_{l}|^2}} +  \frac{v_{l}^*\hat{O}(r) }{ \sqrt{|u_{l}|^2+|v_{l}|^2}}\right)\\
&= \Xi_l(r)\left(\cos(\kappa)\hat{O}^{\dagger}(r)  +  \sin(\kappa)e^{i\theta_\mathrm{s}}\hat{O}(r) \right)
\end{align}
%\textcolor{red}{In the main text, we have only considered coupling between the dipole transition to the 
%quadratures of the electromagnetic vacuum. However, understanding the effect of squeezing
%on the Jaynes-Cummings type coupling between the dipoles and the electromagnetic fluctuations.
%The interaction Hamiltonian for Jaynes-Cummings coupling is given by 
%\begin{align}
%H_{\mathrm{JC}} = \sum_{\mathbf{l}} \hat{O}^{\dagger}a_{\mathbf{l}} + h.c.  
%\end{align}
%Applying the squeezing transformation to the bath operators, we obtain
%\begin{align}
%\nonumber \tilde{H}_{\mathrm{JC}} &= \sum_{\mathbf{l}} \eta_{\mathbf{l}} \hat{O}^{\dagger}(u_{\mathbf{l}}a_{\mathbf{l}}+v_{\mathbf{l}}a^{\dagger}_{-\mathbf{l}}) + h.c.\\\
%\nonumber& =  \sum_{\mathbf{l}}\eta_{\mathbf{l}}  a_{\mathbf{l}}(u_{\mathbf{l}}\hat{O}^{\dagger} + v^*_{\mathbf{l}}\hat{O}^{\dagger} )+ h.c.\\
%&\equiv \sum_{\mathbf{l}} \eta_{\mathbf{l}} a_{\mathbf{l}}\hat{\tilde{O}}^{\dagger}_{\mathbf{l}}+ h.c.
%\end{align}}
Then, assuming that the dipole frequency is much smaller than the squeezed vacuum frequencies,
the effective interactions mediated by the squeezed vacuum becomes
\begin{widetext}
\begin{align}
\nonumber H_{\mathrm{int}}(r,r') &= (\Xi^*(r)\Xi(r'))\sum_{l}(|u_{l}|^2+|v_{l}|^2)\eta^2_{l}/\omega_l \tau^{\dagger}_{l}(r)\tau_{l}(r') + h.c.\\
\nonumber&= (\Xi^*(r)\Xi(r')) \sum_{l}(|u_{l}|^2+|v_{l}|^2)\frac{\eta^2_{l}}{\omega_l} \bigg\{\hat{O}^{\dagger}(r)\hat{O}(r')+\hat{O}^{\dagger}(r')\hat{O}(r)\\
&+2\cos(\kappa)\sin(\kappa)\cos(\theta_s) (\hat{O}^{\dagger}(r)\hat{O}^{\dagger}(r')+\hat{O}(r')\hat{O}(r))\bigg\}
\end{align}
\end{widetext}
where $\omega_l$ is the frequency of the squeezed vacuum modes, and we have assumed that
the squeezed vacuum modes have the same phase between the two spatial points in question (i.e., $\arg{(\Xi(r))} = \arg{(\Xi(r'))}$). 
Then, setting the squeezing angle to be $\theta_s = \pi/2$, we obtain the same effective Hamiltonian
as in the case of quadrature coupling, but with the effective interaction strength reduced by half
in the limit of large squeezing 
\begin{align}
H_{\mathrm{int}}(r,r') &\approx  \Xi^*(r)\Xi(r') \sum_{l}e^{2s(\omega_l)}\frac{\tilde{g}_{\mathrm{l}}}{2}\hat{O}^{\dagger}(r)\hat{O}(r')+h.c..
\end{align}
Note that if we identify the dipole operator with density $\hat{O}(r) = \psi^{\dagger}(r)\psi(r)$,
or any other Hermitian operator, we again recover an enhancement of $e^{2s(\omega_l)}$ for $\theta_s=0$.
Thus, the squeezed vacuum also enhances interactions of JC type. 
%On the other hand when $\hat{O}(r)$ is not hermitian, the enhancement is reduced to half 
%in the limit of large squeezing. For $\hat{O}(r) = \sigma^{+}$, JC type coupling results in 
%anisotropy in the transverse exchange interactions. The anisotropy can be fixed by
%implementing two squeezed vacuum beams which are squeezed in orthogonal quadratures
%(i.e., $v_{-l}^{*} \rightarrow -v_{-l}^{*}$ ).
%Generically, when such two beams are used, we obtain
%\begin{align}
%H_{\mathrm{int}} = \sum_{l}2\eta^2_{l}/\omega_l (|u_{l}|^2 + |v_{l}|^2)\hat{O}^{\dagger}\hat{O},
%\end{align}
%where $2(|u_{l}|^2 + |v_{l}|^2)\approx e^{2s(\omega_l)}$ in the limit of large squeezing. 

\section{Comparison to cQED experiments with cold atoms, and dissipative coupling}
\label{app:compcQED}
In this Appendix, we present an order of magnitude comparison for the coupling strength in the squeezed vacuum 
setup and the recently realised cQED setup where a supersolid transition was observed. We show that by a combination
vacuum squeezing and the modification of the transverse mode profile of the squeezed vacuum beam, a supersolid transition
can be realised in the squeezed vacuum setup. 

The supersolid transition using squeezed vacuum is feasible if the enhancement of the dipole interactions due to
vacuum squeezing matches the enhancement due to the cavity confinement in the recent cQED experiments \cite{Baumann2010}.
First, we note that the enhancement of the effective interactions due to the cavity
confinement is given by twice the finesse $\mathcal{F}$,
\begin{align}
\mathrm{max}\left(\frac{\tilde{g}_{cQED}}{\tilde{g}_{FS}}\right) \approx \frac{2\pi c }{\kappa_c L_{cav}} = 2 \mathcal{F},
\label{eq:finesse}
\end{align}
where we have assumed that the transverse cross-sectional areas of the cavity and 
the cavity mode are the same for simplicity. In Eq. ($\ref{eq:finesse}$),  $\kappa_c$ is the cavity
decay rate which sets the smallest detuning between the atomic transition and the cavity 
resonance that one can achieve. On the other hand, $2\pi c/ L_{cav}$ is the density of states per angular frequency
within the cavity volume. That is, the effective interactions are enhanced by number of round trips 
that a single photon does before it leaves the cavity. 

On the other hand, the squeezed vacuum has two factors that contribute 
to the enhancement of the dipole coupling strength. Besides the aforementioned degree of squeezing which increases 
the fluctuations in a given vacuum quadrature, the finite squeezing bandwidth contributes to the enhancement of the 
effective interactions by increasing the number of vacuum modes that couple to the atomic degrees of freedom. 
However, a constant squeezing of 20 dB over a bandwidth of $10$ MHz, the enhancement
of the vacuum coupling due to squeezing is at best $1\%$ of the enhancement due to the optical 
cavity as we have shown in the main text.

However, for \textit{collective} effects demonstrated in the cQED setup, another two orders of magnitude enhancement
can be achieved by increasing the density of particles in the squeezed vacuum mode. Here, the spatial tunability
of the squeezed vacuum modes becomes useful. As discussed in the Introduction,  in cQED, the spatial profiles of 
the vacuum modes are determined by the reflective walls of the cavity. Usually, transverse modes of a stable cavity are described 
by Laguerre-Gaussian(LG) functions \cite{siegman1986lasers}. When an atomic cloud is placed inside the Laguerre-Gaussian 
mode, different regions of the cloud see a spatially inhomogeneous vacuum amplitude unless the size of the 
cloud is much smaller than the diameter of the LG mode. Because such an inhomogeneity
can cause unwanted and uncontrolled complications, the size of the atomic system is restricted to a fraction of the cavity mode width.
On the other hand, the squeezed vacuum mode can be tailored to have a constant amplitude over the whole
cross-sectional area of the squeezed vacuum beam. Therefore, the size of the atomic cloud can be made as large 
as the transverse area of the squeezed vacuum mode. Hence with the additional collective enhancement due to the density
of atoms inside the vacuum mode, the resulting effective interaction strength becomes comparable to the that recently 
demonstrated in recent cQED  \cite{Baumann2010}.

\section{The derivation of the model in Eq. ($\ref{eq:BosHam}$) for $d = 1$}
\label{app:BoseEx}
Here we give a short derivation of the model in Eq. ($\ref{eq:BosHam}$), where the 
stimulated Raman transitions take the atom back to its initial state. We follow closely the derivation 
presented in Ref. \cite{Ritsch2013}. This example corresponds to the 
system realized in the recent cold atom experiments in Ref. \cite{Baumann2010}, and is a way to 
realize the coupling to the atomic dipole operator in Eq. ($\ref{eq:BEC}$). We start by writing the 
full Hamiltonian of the system in the rotating frame of the pump $P$ whose spatially varying 
coupling strength is $\Omega_{\mathrm{p}}(r)\in \mathbf{R}$

\begin{align}
\nonumber H &= H_A+H_R+H_{A-A}+H_{A-P}+H_{A-R}\\
\nonumber H_A&= \int d^2 r \Psi_g^{\dagger}(r)\left(-\frac{\hbar^2}{2 m}\nabla^2 \right)\Psi_g(r)\\
\nonumber & + \Psi^{\dagger}_e(r)\left(-\frac{\hbar^2}{2 m}\nabla^2 -\Delta_a \right)\Psi_e(r)\\
\nonumber H_R&= \sum_{l} \int d^2r  \hbar \omega_l |\tilde{\Xi}_{l} |^2 (r) \bar{a}_l^{\dagger}\bar{a}_l\\
\nonumber H_{A-A}&=  \frac{U}{2}\int d^2r \Psi_g^{\dagger}(r) \Psi_g^{\dagger}(r) \Psi_g(r) \Psi_g(r) \\
\nonumber H_{A-P}&=  -i \hbar \int d^2 r \Psi^{\dagger}_g(r) \Psi_e(r)\Omega_{\mathrm{p}}(r)+h.c\\
\nonumber  H_{A-R}&=-i \hbar\sum_{l} \int d^2r \Psi_g(r)^{\dagger} \Psi_e(r)\bar{a}_{l}^{\dagger}e^{s(\omega_l)}g_l(r) + h.c,
\end{align}
where $H_{A(R)}$ are the atom (reservoir) Hamiltonians, and $H_{X-Y}$ describe the interaction between the
$X$ and $Y$. $\Delta_{a}$ is the detuning between the pump and the atomic transition to the excited state, $U$
is the strength of the contact interaction between the atoms, and $g_l(r) \equiv g_l \tilde{\Xi}_{l}(r)$, with $g_l$ is the 
dipole coupling strength to mode $l$.

In the steady state, the excited state can be adiabatically eliminated 
given that the atom - pump detuning $\Delta_a$ is much larger than the $\Omega_p$ and $g_l$.
To this end, we write,
\begin{align}
\Psi_e(x) \approx -\frac{i}{\Delta_a}\left(\sum_l g_l(r)\bar{a}_l + \Omega_{\mathrm{p}}(r)\right)\Psi_g(x).
\label{excitedSS}
\end{align}
We can eliminate the excited state by plugging in the above solution for $\Psi_{e}(x)$ to the equation of motion of the 
ground state. The resulting evolution of the ground state is generated by the effective Hamiltonian
\begin{align}
\nonumber H_{eff}  &= \int d^{2}r \Psi_g^{\dagger}(r)\bigg\{ -\frac{\hbar^2}{2 m} \nabla^2 + \frac{U}{2}\int d^2 r \Psi_g^{\dagger}(r)\Psi_g(r) \\
\nonumber &+ \frac{\hbar}{\Delta_a}\left(|\Omega_{\mathrm{p}}|^2(r)+ \sum_{l,l'}g_{l}^*(r)g_{l'}(r)\bar{a}_{l}^{\dagger}\bar{a}_{l'}\right)\\
\nonumber &  +\frac{\hbar}{\Delta_a}\left(\sum_{l}\Omega_{\mathrm{p}}g_{l}(r)(\bar{a}_l + \bar{a}^{\dagger}_{l})\right)\bigg\} \Psi_g(r)\\
 & + \sum_{l} \hbar\omega_l |\tilde{\Xi}_{l} |^2 \bar{a}_l^{\dagger}\bar{a}_l,
\label{eq:EffHam}
\end{align}
where we assumed that $g_{l}(r) \in \mathbf{R}$. Importantly, the form of the interaction in the third line of Eq. ($\ref{eq:EffHam}$)
is the same as the Hamiltonian in Eq. ($\ref{eq:BosHam}$), with the identification $\eta_l = \Omega_{\mathrm{p}}(r) g_{l}(r)/ \Delta_a$
and $\hat{O} = \Psi_g^{\dagger}\Psi_g$. 

%We are interested in the effective coupling between the squeezed vacuum modes and the Bogolyubov excitations on top 
%of an interacting condensate whose amplitude is given by $\alpha \in \mathbf{R}$. To this end, we displace the atomic ground state field as 
%\begin{align}
%\Psi_{g}(r) \rightarrow  \alpha + \delta \Psi_{g}(r), 
%\end{align}
%such that $\delta \Psi_{g}(r)$ describe the quantum fluctuations on top of the condensate.
%Thus, we obtain the quadrature coupling term between the quantum fluctuations on top of the 
%atomic condensate and those on top of the squeezed vacuum
%\begin{align}
%H_{\mathrm{int}} = \hbar \alpha \sum_{l}\frac{\Omega_{\mathrm{p}}(r) g_{l}(r)}{\Delta_a}(\bar{a}_l +\bar{a}^{\dagger}_{l})[\delta\Psi_g(r)+\delta\Psi^{\dagger}_g(r)].
%\end{align}
%
We note that a single coherent drive was enough to implement the quadrature coupling interaction 
because the stimulated Raman transition brings the atom back to its ground state $g$. When this is 
not the case, at least two coherent sources are necessary.

%
%\bibliography{refpaper.bib}

\begin{thebibliography}{10}

\bibitem{nielsen2010quantum}
M.A. Nielsen and I.L. Chuang.
\newblock {\em Quantum Computation and Quantum Information: 10th Anniversary
  Edition}.
\newblock Cambridge University Press, 2010.

\bibitem{Cirac1995}
J.~I. Cirac and P.~Zoller.
\newblock Quantum computations with cold trapped ions.
\newblock {\em Phys. Rev. Lett.}, 74:4091--4094, May 1995.

\bibitem{Wallraff2004}
A.~Wallraff, D.~I. Schuster, A.~Blais, L.~Frunzio, R.~S. Huang, J.~Majer,
  S.~Kumar, S.~M. Girvin, and R.~J. Schoelkopf.
\newblock Strong coupling of a single photon to a superconducting qubit using
  circuit quantum electrodynamics.
\newblock {\em Nature}, 431(7005):162--167, 09 2004.

\bibitem{Majer2007}
J.~Majer, J.~M. Chow, J.~M. Gambetta, Jens Koch, B.~R. Johnson, J.~A. Schreier,
  L.~Frunzio, D.~I. Schuster, A.~A. Houck, A.~Wallraff, A.~Blais, M.~H.
  Devoret, S.~M. Girvin, and R.~J. Schoelkopf.
\newblock Coupling superconducting qubits via a cavity bus.
\newblock {\em Nature}, 449(7161):443--447, 09 2007.

\bibitem{Imamoglu1999}
A.~Imamoglu, D.~D. Awschalom, G.~Burkard, D.~P. DiVincenzo, D.~Loss,
  M.~Sherwin, and A.~Small.
\newblock Quantum information processing using quantum dot spins and cavity
  qed.
\newblock {\em Phys. Rev. Lett.}, 83:4204--4207, Nov 1999.

\bibitem{Saffman2010}
M.~Saffman, T.~G. Walker, and K.~M\o{}lmer.
\newblock Quantum information with rydberg atoms.
\newblock {\em Rev. Mod. Phys.}, 82:2313--2363, Aug 2010.

\bibitem{Blatt2012}
R.~Blatt and C.~F. Roos.
\newblock Quantum simulations with trapped ions.
\newblock {\em Nat Phys}, 8(4):277--284, 04 2012.

\bibitem{Landig2016}
Renate Landig, Lorenz Hruby, Nishant Dogra, Manuele Landini, Rafael Mottl,
  Tobias Donner, and Tilman Esslinger.
\newblock Quantum phases from competing short- and long-range interactions in
  an optical lattice.
\newblock {\em Nature}, 532(7600):476--479, 04 2016.

\bibitem{Schreiber2016}
Michael Schreiber, Sean~S. Hodgman, Pranjal Bordia, Henrik~P. L{\"u}schen,
  Mark~H. Fischer, Ronen Vosk, Ehud Altman, Ulrich Schneider, and Immanuel
  Bloch.
\newblock Observation of many-body localization of interacting fermions in a
  quasirandom optical lattice.
\newblock {\em Science}, 349(6250):842--845, 2015.

\bibitem{Labuhn2016}
Henning Labuhn, Daniel Barredo, Sylvain Ravets, Sylvain de~L{\'e}s{\'e}leuc,
  Tommaso Macr{\`\i}, Thierry Lahaye, and Antoine Browaeys.
\newblock Tunable two-dimensional arrays of single rydberg atoms for realizing
  quantum ising models.
\newblock {\em Nature}, advance online publication:--, 06 2016.

\bibitem{Haroche2006exploring}
S.~Haroche and J.M. Raimond.
\newblock {\em Exploring the Quantum: Atoms, Cavities, and Photons}.
\newblock Oxford Graduate Texts. OUP Oxford, 2006.

\bibitem{Purcell1946}
E.~M. Purcell, H.~C. Torrey, and R.~V. Pound.
\newblock Resonance absorption by nuclear magnetic moments in a solid.
\newblock {\em Phys. Rev.}, 69:37--38, Jan 1946.

\bibitem{Lamb1947}
Willis~E. Lamb and Robert~C. Retherford.
\newblock Fine structure of the hydrogen atom by a microwave method.
\newblock {\em Phys. Rev.}, 72:241--243, Aug 1947.

\bibitem{gardiner2004quantum}
C.~Gardiner and P.~Zoller.
\newblock {\em Quantum Noise: A Handbook of Markovian and Non-Markovian Quantum
  Stochastic Methods with Applications to Quantum Optics}.
\newblock Springer Series in Synergetics. Springer, 2004.

\bibitem{Hakioglu1997}
T.~Hakio\ifmmode~\breve{g}\else \u{g}\fi{}lu and H.~T\"ureci.
\newblock Correlated phonons and the ${T}_{c}$-dependent dynamical phonon
  anomalies.
\newblock {\em Phys. Rev. B}, 56:11174--11183, Nov 1997.

\bibitem{Cotlet2016}
Ovidiu Cotle\ifmmode~\mbox{\c{t}}\else \c{t}\fi{}, Sina
  Zeytino\ifmmode~\check{g}\else \v{g}\fi{}lu, Manfred Sigrist, Eugene Demler,
  and Ata\c{c} Imamo\ifmmode~\check{g}\else
  \v{g}\fi{}lu.
\newblock Superconductivity and other collective phenomena in a hybrid
  bose-fermi mixture formed by a polariton condensate and an electron system in
  two dimensions.
\newblock {\em Phys. Rev. B}, 93:054510, Feb 2016.

\bibitem{Knap2016}
Knap, Michael and Babadi, Mehrtash and Refael, Gil and Martin, Ivar and Demler, Eugene.
\newblock Dynamical Cooper pairing in non-equilibrium electron-phonon systems.
\newblock {\em Phys. Rev. B}, 94:214504, Dec 2016.

\bibitem{Lemonde2016}
Marc-Antoine Lemonde, Nicolas Didier, and Aashish~A. Clerk.
\newblock Enhanced nonlinear interactions in quantum optomechanics via
  mechanical amplification.
\newblock {\em Nature Communications}, 7:11338 EP --, 04 2016.

\bibitem{Lu2015}
Xin-You L\"u, Ying Wu, J.~R. Johansson, Hui Jing, Jing Zhang, and Franco Nori.
\newblock Squeezed optomechanics with phase-matched amplification and
  dissipation.
\newblock {\em Phys. Rev. Lett.}, 114:093602, Mar 2015.

\bibitem{Mehmet2011}
Moritz Mehmet, Stefan Ast, Tobias Eberle, Sebastian Steinlechner, Henning
  Vahlbruch, and Roman Schnabel.
\newblock Squeezed light at 1550 nm with a quantum noise reduction of 12.3 db.
\newblock {\em Opt. Express}, 19(25):25763--25772, Dec 2011.

\bibitem{Gardiner1986}
C.~W. Gardiner.
\newblock Inhibition of atomic phase decays by squeezed light: A direct effect
  of squeezing.
\newblock {\em Phys. Rev. Lett.}, 56:1917--1920, May 1986.

\bibitem{Carmichael1987}
H.~J. Carmichael, A.~S. Lane, and D.~F. Walls.
\newblock Resonance fluorescence from an atom in a squeezed vacuum.
\newblock {\em Phys. Rev. Lett.}, 58:2539--2542, Jun 1987.

\bibitem{Walls1983}
D.~F. Walls.
\newblock Squeezed states of light.
\newblock {\em Nature}, 306(5939):141--146, 11 1983.

\bibitem{Dalton1999}
B.~J. Dalton, Z.~Ficek, and S.~Swain.
\newblock Atoms in squeezed light fields.
\newblock {\em Journal of Modern Optics}, 46(3):379--474, 1999.

\bibitem{Cirac2012}
J.~Ignacio Cirac and Peter Zoller.
\newblock Goals and opportunities in quantum simulation.
\newblock {\em Nat Phys}, 8(4):264--266, 04 2012.

\bibitem{Tomaru2006}
Tatsuya Tomaru and Masashi Ban.
\newblock Secure optical communication using antisqueezing.
\newblock {\em Phys. Rev. A}, 74:032312, Sep 2006.

\bibitem{Bloembergen1996nonlinear}
N.~Bloembergen.
\newblock {\em Nonlinear Optics}.
\newblock W.A. Benjamin, New York, 1995.

\bibitem{Sachdev1984}
Subir Sachdev.
\newblock Atom in a damped cavity.
\newblock {\em Phys. Rev. A}, 29:2627--2633, May 1984.

\bibitem{Carmichael1999statistical}
H.~Carmichael.
\newblock {\em Statistical Methods in Quantum Optics 1: Master Equations and
  Fokker-Planck Equations}.
\newblock Physics and Astronomy Online Library. Springer, 1999.

\bibitem{Note1}
The enhancement of vacuum fluctuations is expressed in decibels as $e^{(2s)}=
  10^{X\protect \mathrm {dB}/10}$. Thus, 20 dB anti-squeezing gives and
  enhancement of factor 100 in the fluctuations.

\bibitem{Vahlbruch2008}
Henning Vahlbruch, Moritz Mehmet, Simon Chelkowski, Boris Hage, Alexander
  Franzen, Nico Lastzka, Stefan Go\ss{}ler, Karsten Danzmann, and Roman
  Schnabel.
\newblock Observation of squeezed light with 10-db quantum-noise reduction.
\newblock {\em Phys. Rev. Lett.}, 100:033602, Jan 2008.

\bibitem{Ritsch2013}
Helmut Ritsch, Peter Domokos, Ferdinand Brennecke, and Tilman Esslinger.
\newblock Cold atoms in cavity-generated dynamical optical potentials.
\newblock {\em Rev. Mod. Phys.}, 85:553--601, Apr 2013.

\bibitem{Dimer2007}
F.~Dimer, B.~Estienne, A.~S. Parkins, and H.~J. Carmichael.
\newblock Proposed realization of the dicke-model quantum phase transition in
  an optical cavity qed system.
\newblock {\em Phys. Rev. A}, 75:013804, Jan 2007.

\bibitem{Buchhold2013}
Michael Buchhold, Philipp Strack, Subir Sachdev, and Sebastian Diehl.
\newblock Dicke-model quantum spin and photon glass in optical cavities:
  Nonequilibrium theory and experimental signatures.
\newblock {\em Phys. Rev. A}, 87:063622, Jun 2013.

\bibitem{Douglas2015}
Douglas J. S., Habibian H., Hung C. L., Gorshkov A. V., Kimble H. J., and Chang D. E.
\newblock Quantum many-body models with cold atoms coupled to photonic
  crystals.
\newblock {\em Nat Photon}, 9(5):326--331, 05 2015.

\bibitem{Strack2011}
Philipp Strack and Subir Sachdev.
\newblock Dicke quantum spin glass of atoms and photons.
\newblock {\em Phys. Rev. Lett.}, 107:277202, Dec 2011.

\bibitem{Gopalakrishnan2011}
Sarang Gopalakrishnan, Benjamin~L. Lev, and Paul~M. Goldbart.
\newblock Frustration and glassiness in spin models with cavity-mediated
  interactions.
\newblock {\em Phys. Rev. Lett.}, 107:277201, Dec 2011.

\bibitem{Note2}
These terms can become relevant if a coherent state spontaneously builds up in
  the vacuum modes. For instance, such an effect is responsible for the chaotic
  behaviour of the single mode Dicke model near the phase transition. However,
  as we are working with free space squeezed modes, the number of
  electromagnetic vacuum modes is large enough to suppress the chaotic
  behaviour \cite {Tolkunov2007}.

\bibitem{Note3}
We note that the enhanced fluctuations in the anti squeezed quadrature does not
  pose a problem for the detection scheme when the polarization of the squeezed
  vacuum modes are used appropriately. The light emitted from the atom into the
  squeezed vacuum can be measured independently from the squeezed vacuum by
  using the polarization of the squeezed vacuum modes \cite {Delteil2016}. When
  the polarisation of the squeezed vacuum beam is not aligned with the
  polarisation of the light emitted by the atoms, the emitted light can be
  separated from the fluctuations in the anti-squeezed quadrature of the vacuum
  by the use of a polariser.

\bibitem{Kraus2004}
B.~Kraus and J.~I. Cirac.
\newblock Discrete entanglement distribution with squeezed light.
\newblock {\em Phys. Rev. Lett.}, 92:013602, Jan 2004.

\bibitem{Goldstein1996}
V. Goldstein and P. Meystre.
\newblock Dipole-dipole interaction in squeezed vacua.
\newblock {\em Phys. Rev. A}, A 53, 3573 , 1996.

\bibitem{Banarjee2010}
Subhashish Banerjee and V. Ravishankar and R. Srikanth.
\newblock Dynamics of entanglement in two-qubit open system interacting with a squeezed thermal bath via dissipative interaction.
\newblock {\em Annals of Physics},  325(4) 816, 2010.

\bibitem{Lodahl2017}
Peter Lodahl, Sahand Mahmoodian, S{\o}ren Stobbe, Arno Rauschenbeutel, Philipp
  Schneeweiss, J{\"u}rgen Volz, Hannes Pichler, and Peter Zoller.
\newblock Chiral quantum optics.
\newblock {\em Nature}, 541(7638):473--480, 01 2017.

\bibitem{Ramos2014}
Tom\'as Ramos, Hannes Pichler, Andrew~J. Daley, and Peter Zoller.
\newblock Quantum spin dimers from chiral dissipation in cold-atom chains.
\newblock {\em Phys. Rev. Lett.}, 113:237203, Dec 2014.

\bibitem{Kulkarni2013}
Manas Kulkarni, Baris \"Oztop, and Hakan~E. T\"ureci.
\newblock Cavity-mediated near-critical dissipative dynamics of a driven
  condensate.
\newblock {\em Phys. Rev. Lett.}, 111:220408, Nov 2013.

\bibitem{Kim1989}
M.~S. Kim, F.~A.~M. de~Oliveira, and P.~L. Knight.
\newblock Properties of squeezed number states and squeezed thermal states.
\newblock {\em Phys. Rev. A}, 40:2494--2503, Sep 1989.

\bibitem{goodman2007speckle}
J.W. Goodman.
\newblock {\em Speckle Phenomena in Optics: Theory and Applications}.
\newblock Roberts \& Company, 2007.

\bibitem{Haeberlen1968}
U.~Haeberlen and J.~S. Waugh.
\newblock Coherent averaging effects in magnetic resonance.
\newblock {\em Phys. Rev.}, 175:453--467, Nov 1968.

\bibitem{Tanji2011}
Haruka Tanji-Suzuki, Ian~D Leroux, Monika~H Schleier-Smith, Marko Cetina,
  Andrew~T Grier, Jonathan Simon, and Vladan Vuletic.
\newblock Interaction between atomic ensembles and optical resonators:
  classical description.
\newblock {\em arXiv preprint arXiv:1104.3594}, 2011.

\bibitem{loudon1983quantum}
R.~Loudon.
\newblock {\em The quantum theory of light}.
\newblock Oxford science publications. Clarendon Press, 1983.

\bibitem{Semmler2016}
Marion Semmler, Stefan Berg-Johansen, Vanessa Chille, Christian Gabriel, Peter
  Banzer, Andrea Aiello, Christoph Marquardt, and Gerd Leuchs.
\newblock Single-mode squeezing in arbitrary spatial modes.
\newblock {\em Opt. Express}, 24(7):7633--7642, Apr 2016.

\bibitem{Prokofiev2012}
Massimo Boninsegni and Nikolay~V. Prokof'ev.
\newblock \textit{Colloquium} : Supersolids: What and where are they?
\newblock {\em Rev. Mod. Phys.}, 84:759--776, May 2012.

\bibitem{Prokofiev2007}
Nikolay Prokof'ev.
\newblock What makes a crystal supersolid?
\newblock {\em Advances in Physics}, 56(2):381--402, 2007.

\bibitem{Wrachtrup2006}
J{\"o}rg Wrachtrup and Fedor Jelezko.
\newblock Processing quantum information in diamond.
\newblock {\em Journal of Physics: Condensed Matter}, 18(21):S807, 2006.

\bibitem{Gardiner1993}
C.~W. Gardiner.
\newblock Driving a quantum system with the output field from another driven
  quantum system.
\newblock {\em Phys. Rev. Lett.}, 70:2269--2272, Apr 1993.

\bibitem{Carmichael1993}
H.~J. Carmichael.
\newblock Quantum trajectory theory for cascaded open systems.
\newblock {\em Phys. Rev. Lett.}, 70:2273--2276, Apr 1993.

\bibitem{cohen1992atom}
C.~Cohen-Tannoudji, J.~Dupont-Roc, and G.~Grynberg.
\newblock {\em Atom-photon interactions: basic processes and applications}.
\newblock Wiley-Interscience publication. J. Wiley, 1992.

\bibitem{mahan2000many}
G.D. Mahan.
\newblock {\em Many-Particle Physics}.
\newblock Physics of Solids and Liquids. Springer, 2000.

\bibitem{Baumann2010}
Kristian Baumann, Christine Guerlin, Ferdinand Brennecke, and Tilman Esslinger.
\newblock Dicke quantum phase transition with a superfluid gas in an optical
  cavity.
\newblock {\em Nature}, 464(7293):1301--1306, 04 2010.

\bibitem{siegman1986lasers}
A.E. Siegman.
\newblock {\em Lasers}.
\newblock University Science Books, 1986.

\bibitem{Tolkunov2007}
Denis Tolkunov and Dmitry Solenov.
\newblock Quantum phase transition in the multimode dicke model.
\newblock {\em Phys. Rev. B}, 75:024402, Jan 2007.

\bibitem{Delteil2016}
Aymeric Delteil, Zhe Sun, Wei-bo Gao, Emre Togan, Stefan Faelt, and Atac
  Imamo{\u g}lu.
\newblock Generation of heralded entanglement between distant hole spins.
\newblock {\em Nat Phys}, 12(3):218--223, 03 2016.


\end{thebibliography}
%\bibliographystyle{unsrt}

%%TC:endignore 
%\end{document}

\bibliographystyle{unsrt}

%TC:endignore 
\end{document}